\documentclass[aps,jmp,amsmath,amssymb,reprint]{revtex4-1}
\usepackage{graphicx}
\usepackage{dcolumn}
\usepackage{bm}
\usepackage{graphicx}
\usepackage{subfigure}
\usepackage{physics}
\usepackage[english]{babel}
\usepackage[usenames, dvipsnames]{color}
\usepackage{hyperref}
\usepackage[normalem]{ulem}
\usepackage[utf8]{inputenc}
\usepackage{romannum}
\usepackage{float}

\begin{document}
\bibliographystyle{apsrev.bst}
\pagenumbering{arabic}
\renewcommand{\figurename}{FIG.}
\def\tablename{TABLE}

\title{Advancing carrier transport models for InAs/GaSb type-II superlattice MWIR photodetectors}        
	\author{Rohit Kumar, Anup Kumar Mandia, Anuja Singh and Bhaskaran Muralidharan}
	\thanks{corresponding author: bm@ee.iitb.ac.in}
	
	\affiliation{Department of Electrical Engineering, Indian Institute of Technology Bombay, Powai, Mumbai-400076, India}
\date{\today}
\begin{abstract}
\noindent In order to provide the best possible performance, modern infrared photodetector designs necessitate extremely precise modeling of the superlattice absorber region. 
We advance the Rode's method for the Boltzmann transport equation in conjunction with the $\bf k.p$ band structure and the envelope function approximation for a detailed computation of the carrier mobility and conductivity of layered type-II superlattice structures, using which, we unravel two crucial insights. First, the significance of both elastic and inelastic scattering mechanisms, particularly the influence of the interface roughness and polar optical phonon scattering mechanisms in technologically relevant superlattice structures. Second, that the structure-specific Hall mobility and Hall scattering factor reveals that temperature and carrier concentrations significantly affect the Hall scattering factor, which deviates significantly from unity even for small magnetic fields. This reinforces the caution that should be exercised when employing the Hall scattering factor in experimental estimations of drift mobilities and carrier concentrations. Our research hence offers a comprehensive microscopic understanding of carrier dynamics in such technologically relevant superlattices. Our models also provide highly accurate and precise transport parameters beyond the relaxation time approximation and thereby paving the way to develop physics-based device modules for mid-wavelength infrared photodetectors.
\end{abstract}
\maketitle
\section{Introduction}
Modeling state-of-the-art infrared (IR) photodetectors \cite{smith1987proposal,rogalski2017inas,mukherjee2021carrier,rogalski2000infrared,dehzangi2021band,rogalski2003infrared} require highly accurate transport parameters for developing dark and photocurrent performance projections \cite{dehzangi2021band,le2019simulation,klipstein2021type,wrobel2012dark,gautam2010performance}. Current technologically relevant IR photodetectors use III-V materials such as InAs/GaSb  \cite{SaiHalasz,manyk2018electronic} due to numerous advantages \cite{rogalski2019type, chow1991type}. Type-II superlattices (T2SLs) based on stacks of InAs/GaSb \cite{chow1991type,rogalski2017inas,smith1987proposal} are thus extensively used to design high-performance third-generation IR detectors \cite{plis2014inas,martyniuk2014new}. Despite the fact that the mobility of the photogenerated minority carriers has a significant impact on the performance of IR photodetectors, carrier transport in technologically relevant T2SL structures has not as extensively been explored. Recent explorations in this context \cite{safa2013study,safa2015role,safa2015study,safa2015vertical,szmulowicz2011calculation,szmulowicz2011calculation2,szmulowicz2013calculation} which include carrier mobility calculations \cite{bastard1981superlattice}, do not conclusively bring to the fore structure-specific impact of important scattering mechanisms such as Piezoelectric (PZ), polar optical phonon (POP), acoustic deformation potential (ADP) scattering mechanisms and most importantly the interface roughness scattering (IRS). \\
\indent With the necessity to develop a deeper understanding of carrier transport in technologically relevant T2SLs, this work advances an accurate model for transport calculations, 
 wherein, we investigate different scattering limited transport under low-field in InAs/GaSb superlattices (SLs) as a function of free electron carrier concentration, temperature, and SL structural parameters. In our calculations, five primary scattering mechanisms that limit carrier mobility are the ionized impurity (II) \cite{conwell1950theory}, the PZ \cite{zook1964piezoelectric}, the ADP \cite{kaasbjerg2013acoustic}, the POP and the IRS \cite{tsai2020application,wataya1989interface,dharssi1991mobility,dharssi1990interface}.\\
\indent  We advance the Rode's method \cite{rode1970electron,rode1973theory,rode1975low} which goes beyond the relaxation time approximation (RTA) \cite{ashcroft1976solid,mermin1970lindhard}, coupled with band structure calculations via the $\bf{k.p}$ \cite{livneh2012k,klipstein2010operator,ricciardi2020ingaas,qiao2012electronic,klipstein2013k} technique that also includes the strain effect due to lattice mismatch between InAs and GaSb materials \cite{aspnes1983dielectric}. We demonstrate the effect of both the elastic and the inelastic scattering mechanisms \cite{datta2005quantum} on the electron mobility of the composite structure for a wide range of temperatures and doping concentrations. Our studies reveal that the low-temperature mobility of T2SLs is limited by the II, PZ and IRS scattering mechanisms. In contrast, the mobility at higher temperatures is mainly limited by the POP scattering mechanism, an inelastic and anisotropic process. At intermediate temperatures, however, the mobility decreases due to a combined effect of ADP and IRS mechanisms. The effects of several structural parameters including layer thicknesses, interface roughness heights, correlation lengths, and ion densities are thoroughly investigated. 
Our calculations thereby reinforce the superiority of the Rode's method \cite{rode1970electron,rode1975low} over the conventionally employed RTA, wherein, the former is applicable over a wide temperature range in the presence of inelastic and anisotropic scattering mechanism.\\ 
\indent In order to experimentally obtain the carrier concentration and drift mobility in a SL structure, it is also important to ascertain the Hall scattering factor, which is frequently thought of as being equal to unity, indicating that the Hall mobility and the drift mobility are equal. However, in many heterostructures, it differs significantly from unity, which results in inaccurate estimates of the carrier density and drift mobility. We clearly show that the temperature and carrier concentrations significantly affect the Hall scattering factor, and that it ranges from 0.3 to about 1.48 even for weak magnetic fields, thereby reinforcing that caution should be exercised when employing this factor in calculations involving drift mobility and carrier concentration. The models developed here pave the way to develop physics-based device modules for mid-wavelength IR (MWIR) photodetectors.\\
\indent This paper is structured as follows. In Sec. \ref{sec_method} we describe the ${\bf k.p}$ model to compute the band structure, electron distribution function, Boltzmann transport formalism, Rode's approach and various scattering processes. In Sec. \ref{simulation_approach} we illustrate the simulation methodology. In Sec. \ref{sec_result}, we explain the findings and finally, in Sec. \ref{conclu}, we summarize our results.

\begin{figure*}[!htbp]
	\centering
	\subfigure[]{\includegraphics[height=0.25\textwidth,width=0.4\textwidth]{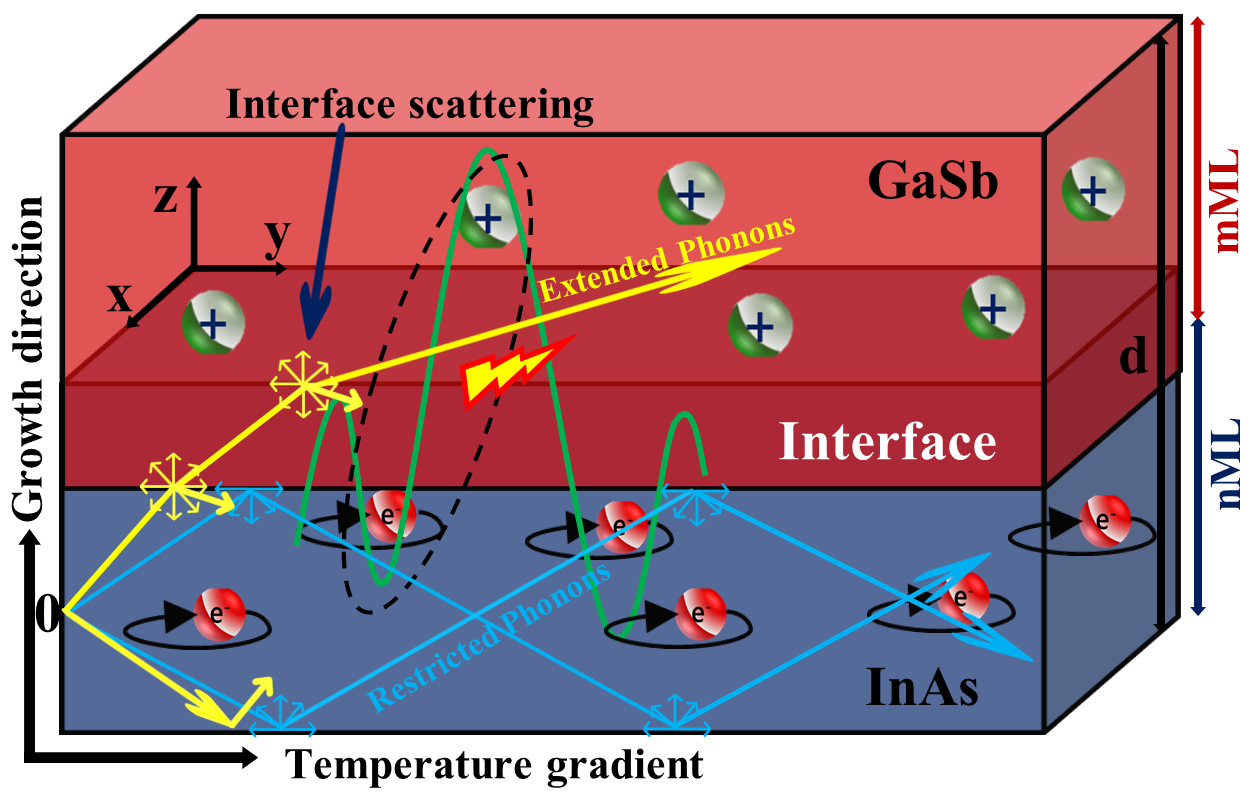}\label{schematic}}
	\quad
	\subfigure[]{\includegraphics[height=0.3\textwidth,width=0.45\textwidth]{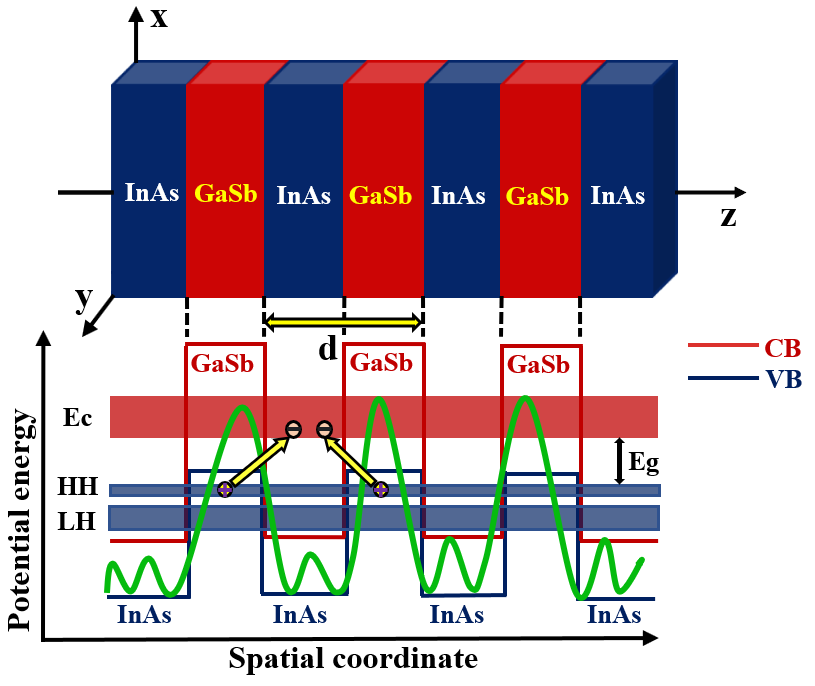}\label{Band_alignment}}
	\quad
	\caption{Preliminaries. (a) Schematic of InAs/GaSb based T2SL structure. The electron wave function in the InAs layer extends beyond the interface into the GaSb layer and overlaps with the heavy hole wave function. Here, nML and mML are the numbers of monolayers of InAs and GaSb respectively, in a single period $d$. (b) Band alignment of InAs/GaSb based T2SL system showing the optical transition between the heavy-hole valence miniband and the electrons from lowest conduction minibands that is employed to detect IR radiation. The periodic potential of the $d$ period emerges in the material due to the modulation of the semiconductor layers. The creation of hole (electron) minibands in the valence (conduction) band is caused by the overlap of hole (electron) wave functions between adjacent GaSb (InAs) layers. The difference between the first electron miniband in the CB and the first heavy hole miniband in the valence band is used to compute the effective bandgap energy $E_g$ of the T2SL (highlighted in black). 
	}
	\label{T2SL structure}
\end{figure*}
\section{Analytical Formalism}
\label{sec_method}
\subsection{Electronic band structure}
\label{k.p}
The energy band structure of T2SLs can be calculated using various theoretical approaches like the density functional theory (DFT) \cite{garwood2017electronic}, the empirical tight-binding method \cite{ashcroft1976solid,wei2004modeling,nucho1978tight}, the empirical pseudopotential method \cite{dente1999pseudopotential,magri2002effects}, many-body perturbation theory \cite{taghipour2018many} and the $\bf{k.p}$ perturbation method \cite{livneh2012k}. For this study, we use the $\bf{k.p}$ technique with the envelope function approximation (EFA) \cite{bastard1981superlattice,bastard1982theoretical,altarelli1983electronic} since it overcomes the computational limitations of first-principles methods. The $\bf{k.p}$ model is extensively used because of its superiority in computing the energy band gap. Unlike $ab-initio$ and tight binding methods, the $\bf{k.p}$ technique requires fewer input parameters \ref{table1}, with the related calculation procedure being straightforward.\\ 
\indent In this work, we solve the 8-band Kane Hamiltonian \cite{kane1980band}, by perturbatively extending the wave function around high-symmetry points of the reciprocal space, employing the Lowdin's perturbation approach \cite{kane1980band}. We also consider the spin-orbit coupling \cite{chuang2012physics} in our computation, which provides additional contributions to the spin splitting of the energy bands \cite{mukherjee2021carrier}. The SL wavefunctions $(\Phi_n(\textbf{z}))$ in the orbital basis states $(u_0(\textbf{z}))$ along the growth direction ($z$) are articulated in terms of the slowly varying envelope functions $(F(\textbf{z}))$, which are given as
\begin{equation}
	\Phi_{n}(\textbf{z}) = \sum_{j}^{}F_{j}(\textbf{z})u_{j0}(\textbf{z}).
	\label{wavefunc}	
\end{equation}
Such envelope functions under the periodic boundary conditions
can be rewritten as
\begin{equation}
	\begin{split}
		F_{j}^{i}(\textbf{k},\textbf{z}_{0}) = e^{-i\textbf{a}d}F_{j}^{i}(\textbf{k},\textbf{z}_{M}),\\
		F_{j}^{i}(\textbf{k},\textbf{z}_{M+1}) = e^{i\textbf{a}d}F_{j}^{i}(\textbf{k},\textbf{z}_{1}),
		\end{split}
	\label{envelope_func}
\end{equation}
where, $d$ denotes the thickness of a period, $M$ represents the number of grid points, $\textbf{a}$ denotes the Bloch vector of the envelope function that spans the Brillouin zone (BZ) and $\textbf{k}$ represents the momentum along the transverse direction. The final Hamiltonian of the SL in the basis set comprises three matrices ($H^{0}$, $H^{\Romannum{1}}$ and $H^{\Romannum{2}}$), given by $H(\textbf{k},k_z)=H^{0}+H^{\Romannum{1}}\left(-i\frac{\partial}{\partial z}\right)+\left(-i\frac{\partial}{\partial z}\right)H^{\Romannum{2}}\left(-i\frac{\partial}{\partial z}\right)$. The entire coupled differential equation is then solved using a numerical finite difference method \cite{jiang2014finite}, as described in earlier work \cite{mukherjee2021carrier}.\\
\indent The interface between the InAs and the GaSb layers is very abrupt as depicted in Fig.~\ref{schematic}. The energy difference between the conduction band minimum (CBM) and the first heavy hole (HH) maximum at the center of the BZ determines the band gap in an InAs/GaSb-based T2SL, as shown in Fig.~\ref{Band_alignment}. Figure \ref{Band_alignment} also demonstrates that the InAs conduction band (CB) is lower than the GaSb valence band (VB), indicating that the band structure is a staggered T2SL \cite{dhar2013advances}.
\subsection{Carrier transport model}
\subsubsection*{\bf\emph{1. \hspace{0.2cm}{Boltzmann transport equation and its solution}}}

\indent In order to characterize the behavior of the T2SL system, we solve the Boltzmann transport equation (BTE) and compute the probability of finding a carrier with a crystal momentum $k$ at a location $r$ at a time $t$ as indicated by the distribution function $f(r,k,t)$. Solving the BTE \eqref{BTE} yields the average distribution of the carriers in both the position and the momentum space. The BTE can be written as \cite{lundstrom2002fundamentals,ferry2016semiconductor,singh2007electronic} 

\begin{equation}
   \frac{\partial f}{\partial t} -\frac{\partial f} {\partial t}\Big|_{_{diff}} -\frac{\partial f} {\partial t}\Big|_{_{forces}} =\frac{\partial f} {\partial t}\Big|_{_{coll}} + s(\bf{r},\bf{p},\emph t) \:.\\ 
\label{BTE}
\end{equation}
The term $s (\bf{r},\bf{p},\emph t)$, in Eq. \eqref{BTE}, represents generation-recombination processes \cite{pierret1987advanced}, where $p$ is the classical momentum. The term $(\partial f/\partial t)_{forces}$, represents the change in the distribution function due to applied electric and magnetic fields. The term $(\partial f/\partial t)_{forces}=-\bf{F} \cdot \nabla_\emph p {\emph f}$, where, {$\bf{F} = (\emph d \bf{p}/\emph {dt}) = \hslash (\emph d\bf{k}/\emph{dt}) = -\emph e (\bf{E} + \bf{v}\times \bf{B}),$} represents the total force equal to the sum of the electric-force and the Lorentz-force owing to the magnetic flux density $\bf{B}$, where $e$ is the electron charge, $\bf{E}$ is the applied electric field and $\bf{v}$ denotes the group velocity of the carriers. The term $(\partial f/\partial t)_{diff}= -\bf{v}. \nabla_\emph r \emph f$, refers to the spatial change in the distribution function caused by temperature or concentration gradients, which results in carrier diffusion in the coordinate space. Here, $(\partial f/\partial t)_{coll}$ is the collision term, which indicates how the distribution function changes over time due to collision events, and can be described as the difference between the in- and the out-scattering processes, i.e.,
\begin{equation}
\begin{split}
    \Big(\frac{\partial f}{\partial t}\Big)_{_{coll}} = {\sum\limits_{\bf{k_{1}}}}\: \Big\{ S(\bf{k_{1},\:k})\:\emph f\:(\bf{k_{1}})\:\Big[ 1-\emph f\:(\bf{k})\Big]\\
    \:-\:S(\bf{k,\:k_{1}})\:\emph f\: (\bf{k})\:\Big[1-\emph f\:(\bf{k_{1}})\Big]\Big\} \: ,
  \label{distribution_func_collision}
  \end{split}
\end{equation} 
\noindent where, {$S(\bf{k},\:{k_{1})}$} and {$S(\bf{k_{1},\:k)}$} are the transition rates for an electron moving between states $\bf{k}$ and $\bf{k_{1}}$. Under steady-state, $\frac{\partial f}{\partial t}=0$, in case of spatial homogeneity, $\nabla_r f=0$, and assuming that there is no recombination-generation term, the BTE \eqref{BTE} can be rewritten as
\begin{equation}
\begin{split}
    -\frac{e\bf{E}}{\hslash} \cdot \nabla_{k}f= {\sum\limits_{\bf{k_{1}}}}\:\Big\{ S(\bf{k_{1},\:k})\:\emph f\:(\bf{k_{1}})\:\Big[1-\emph f\:(\bf{k})\Big]\\
    -\:S(\bf{k,\:k_{1}})\:\emph f\:(\bf{k})\:\Big[1-\emph f\:(\bf{k_{1}})\Big]\Big\} \: .
\end{split}
\label{revised_BTE}
\end{equation} 

In the low-electric field regime, the distribution function can be represented as \cite{vasileska2017computational}

\begin{equation}
     f(\textbf{k})=\emph f_\emph 0\Big[\varepsilon(k)\Big]+\emph g(k)\:\emph{cos}\: \theta\:,\\
    \label{distribution_func}
\end{equation} 
\noindent where, $k$=$|\textbf {k}|$, $f$ denotes the actual electron distribution function, which includes both the elastic and the inelastic scattering mechanisms, $g(k)$ is the perturbation term to $f_0 [\varepsilon (k) ]$ produced by the electric field, $\theta$ is the angle between applied electric field (along the symmetry axis) and the electron wave vector $\bf{k}$, and $f_0$ represents the distribution function under equilibrium conditions, which is taken according to Fermi-Dirac statistics \cite{ashcroft1976solid,pierret1987advanced}. 
\noindent By solving Eqs. \eqref{revised_BTE} and \eqref{distribution_func}, the perturbation term $g(k)$, can be calculated as \cite{rode1970electron,rode1975low,chakrabarty2019semi,mandia2021ammcr}
\begin{equation}
    g_i(k)=\Bigg[\frac{\emph S_\emph i\Big[\emph g_\emph i(k)\Big]-\frac{ (-e){E}}{\hslash}\Big[\frac{\partial \emph f_\emph 0}{\partial k}\Big]}{\emph S_\emph o(k)+\frac{1}{\tau_{el}(k)}}\Bigg] \: ,
    \label{perturbation}
\end{equation} 

\begin{figure}[H]
\centering
\includegraphics[height=0.3\textwidth,width=0.48\textwidth]{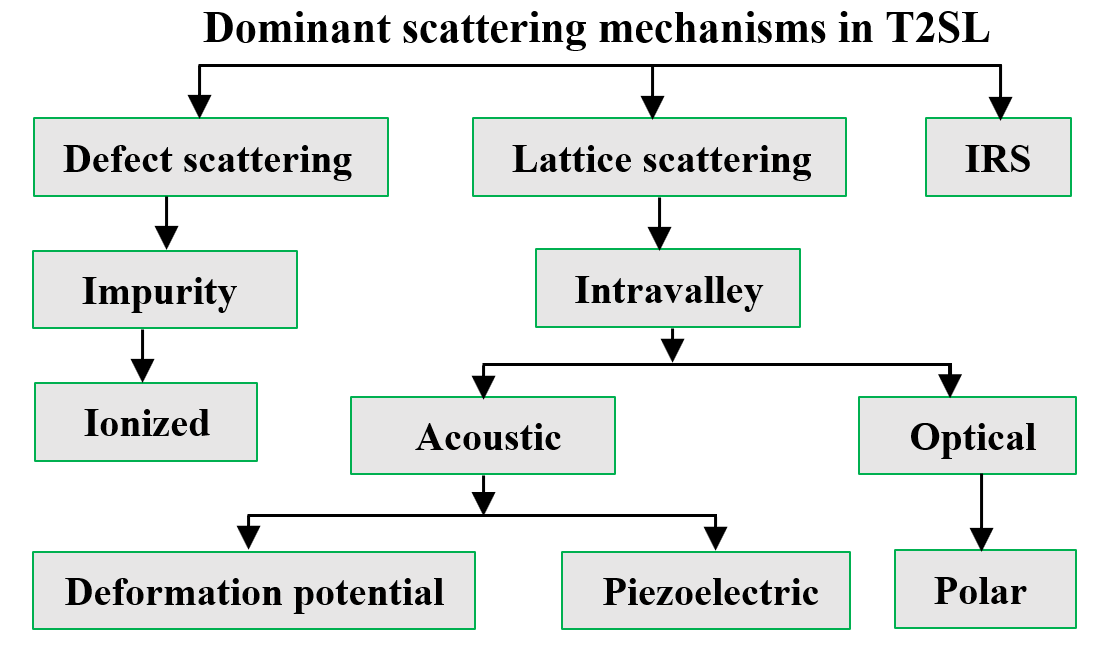}
\caption{ The various dominant scattering mechanisms involved in a T2SL structure.}
\label{fig:dominant_scattering}
\end{figure}

\noindent where $E=|\textbf E|$, and $g_i(k)$ appears on both sides of Eq. \eqref{perturbation}. Hence, we solve Eq. \eqref{perturbation} iteratively and the convergence is exponentially fast which takes a few iterations. Once $g_i(k)$ is obtained, we calculate the mobility. In Eq. \eqref{perturbation}, the term $i$ indicates the iteration index, and the terms, $S_i$ \& $S_o$ are the in-scattering and the out-scattering operators, respectively, for inelastic scattering mechanisms, as explained in Sec. \ref{POP}. The term $\frac{1}{\tau_{el}(k)}$, represents the total momentum relaxation rate of all the elastic scattering mechanisms, which is calculated according to the Matthiessen's rule \eqref{tau_elastic}, and can be written as 

\begin{equation}
    \frac{1}{\tau_{_{el}}(k)}=\frac{1}{\tau_{_{II}}(k)}+\frac{1}{\tau_{_{PZ}}(k)}+\frac{1}{\tau_{_{ADP}}(k)}+\frac{1}{\tau_{_{IRS}}(k)} \: .
    \label{tau_elastic}
\end{equation}

The various dominant scattering mechanisms involved in an InAs/GaSb-based T2SL structure are shown in Fig. \ref{fig:dominant_scattering}.

\subsubsection*{\bf\emph{2. \hspace{0.2cm}{Ionized impurity scattering}}}
\label{II}
The II scattering mechanism \cite{conwell1950theory} arises due to the Coulomb interactions between electrons and ions, when a charged center is introduced inside the bulk material. The II scattering mechanism is entirely elastic and dominates usually at high doping concentrations and low temperatures. The II scattering mechanism dominates near the CB edge but reduces drastically as the energy increases \cite{ganose2021efficient}. The scattering rate for the II increases rapidly with decreasing temperature. Here, we use the Brooks-Herring approach \cite{brooks1955theory} for the calculation of II scattering rate \cite{faghaninia2015ab,rode1975low}, which is given by 
\begin{widetext}
\begin{equation}
    \frac{1}{\tau_{_{II}}(k)}=\frac{e^4N}{8\pi\: {\nu}(k) \:(\epsilon_0\:\epsilon_s)^2\:(\hslash\:k)^2 }\Bigg[P(k)\: ln\Big[1 + 4\:\Big(\frac{k}{\beta}\Big)^2 \Big]-Q(k)\Bigg] \:,
    \label{tau_II}
\end{equation}
\end{widetext}
\noindent where, $\epsilon_0$ is the permittivity of the free space, $\epsilon_s$ is the static dielectric constant, $\hslash$ is the reduced Planck's constant and $N$ is the ionized impurity concentration, which is the sum of the acceptor and donor impurity concentration i.e., $N=N_A+N_D$. Here, $\beta$ indicates the inverse screening length, which is given as
\begin{equation}
    \beta=\sqrt{\frac{e^2}{\epsilon_0\: \epsilon_s\:k_B\:T}\int D_S(\varepsilon)f_0(1-f_0)d\varepsilon} \: ,
    \label{beta}
\end{equation}
\noindent where, $D_S(\varepsilon)$ is the density of states (DOS) at energy $\varepsilon$ and $k_B$ is the Boltzmann constant. $P(k)$ and $Q(k)$ can be expressed as follows \cite{rode1975low,mandia2021ammcr}
\begin{equation}
    P(k)=\Big[\:\frac{3}{4}\Big(\frac{\beta\: c(k)}{k}\Big)^4+2\Big(\frac{\beta\:c(k)}{k}\Big)^2+1\:\Big] \:,
    \label{P}
\end{equation}

\begin{equation}
\begin{split}
    Q(k)=\Big[\frac{3\:\beta^4\: +\: 6\:\beta^2 \: k^2 \:-\:8\:k^4}{(\beta^2\:+\: 4\:k^2)\:k^2}\Big]c^4(k)\\ +\: 8\:\Big[\frac{\beta^2 +2k^2}{\beta^2\:+\:4\:k^2}\Big]\:c^2(k)+\Bigg[\frac{4\:\Big(k/\beta\Big)^2}{1\:+\:4\:\Big(k/\beta\Big)^2}\Bigg] \:.\\
    \end{split}
    \label{Q}
\end{equation}

\noindent The detailed explanation of the $P$ and $Q$ parameters are given in the literature \cite{rode1975low}. Here, the wave function admixture $c(k)$ represents the contribution of the p-orbital to the wave function of the band.

\subsubsection*{\bf\emph{3. \hspace{0.2cm}{Piezoelectric scattering}}}
\label{PZ}
The PZ effect arises due to the acoustic phonon scattering in polar semiconductors. 
Being a weak effect, the PZ scattering is elastic and significant only at low doping concentrations and low temperatures, where other scattering mechanisms are weak. The momentum relaxation rate for the PZ scattering is given by \cite{faghaninia2015ab,rode1970electron}
\begin{equation}
    \frac{1}{\tau_{_{PZ}}(k)}=\frac{(eP)^2\: k_B\:T}{6\pi\: \epsilon_0\:\epsilon_s\: \nu(k)\:\hslash^2}\Big[ 4c^4(k)-6c^2(k)+3 \Big] \: ,\\
    \label{tau_PZ}
\end{equation}
\noindent where, $P$ is a piezoelectric coefficient, which is a dimensionless quantity. 
For the zincblende structure, it is given as \cite{mandia2021ammcr,rode1975low}
\begin{equation}
    P^2=\frac{h_{14}^2\:\epsilon_0\:\epsilon_s\:}{35}\Bigg[{\Big(\frac{12}{c_l}\Big)+\Big(\frac{16}{c_t}\Big)}\Bigg] \:, \\
    \label{piezoelectric_coeff}
\end{equation}
\noindent where, $h_{14}$ is an element of the PZ stress tensor, and $c_t$ and $c_l$ represents the spherically averaged elastic constants for transverse and longitudinal modes, respectively, and are given by \cite{zook1964piezoelectric,rode1975low,rode1970electron}
\begin{equation}
\begin{split}
    c_l=\frac35 c_{11}+\frac15 \Big(2c_{12}+4c_{44}\Big)\:,\\
     c_t=\frac15 \Big(c_{11}-c_{12}\Big)+ \frac35 c_{44} \:,
    \end{split}
    \label{elastic_const}
\end{equation}
where $c_{11}$, $c_{12}$, and $c_{44}$ are three independent elastic constants.
\subsubsection*{\bf\emph{4. \hspace{0.2cm}{Acoustic deformation potential scattering}}}
\label{ADP}
The ADP scattering mechanism is caused by the interaction of electrons with non-polar acoustic phonons. It is approximately elastic near room temperature 
For the ADP scattering mechanism, the momentum relaxation rate is given by \cite{faghaninia2015ab, rode1975low} 
\begin{equation}
    \frac{1}{\tau_{_{ADP}}(k)}=\frac{ k_B\:T\:\Big(e\: \Xi_D \: k\Big)^2}{3\pi\:c_{el}\:{\nu}(k)\:\hslash^2} \Big[6\:c^4(k)-8\:c^2(k)+3\Big] \:,\\
    \label{tau_ADP}
\end{equation}

\noindent where, $c_{el}$ denotes the spherically averaged elastic constant and $\Xi_D$ represents the acoustic deformation potential, which is obtained by the CB shift (in eV) per unit strain, owing to the acoustic waves\eqref{deformation_potential}. To calculate the acoustic deformation potential $(\Xi_D)$, we use the following relation \eqref{deformation_potential}
\begin{equation}
    \Xi_D=-V\times\Bigg(\frac{\partial E_{CBM}}{\partial V}\Bigg)\Bigg|_{_{V=V_0}} \:, \\
    \label{deformation_potential}
\end{equation}
where, $V$ denotes the volume, $E_{CBM}$ represents the energy of the CBM and $V_0$ is the zero pressure volume of the structure.

\subsubsection*{\bf\emph{5. \hspace{0.2cm}Interface roughness scattering}}
\label{IRS}
The existence of the interface roughness in a T2SL \cite{safa2015role,szmulowicz2013calculation,safa2013study,wataya1989interface,kothari2017phonon} structure leads to endemic variations in InAs well widths, causes modulation of 
the associated energy levels and introduces an unstable potential for the motion of the confined electrons. 
The IRS mechanism can occur due to the imperfections that arise during the growth of the material. 
The earlier related works \cite{sakaki1987interface,gold1987electronic} show that the degree of scattering decreases in proportion to the well width hence it is important in MWIR detectors. The IRS mechanism is an elastic process and dominates at low temperatures in thin-film systems for a short period of T2SL, and it is significant at high electron density. The momentum relaxation rate for the IRS mechanism is given as \cite{sang2013dislocation,ferry2016semiconductor,goodnick1985surface}

\begin{equation}
\begin{split}
    \frac{1}{\tau_{_{IRS}}(k)}=\Bigg(\frac{e^2\: \Delta \:\Lambda}{\epsilon_0\:\epsilon_\infty}\Bigg)^{^2}\:\frac{k}{\hslash^2\:\nu(k)}\:
    \Bigg(N_d + \frac{N_s}{2}\Bigg)^2\\
   \times\: \frac{1}{\sqrt{1+(k\Lambda)^2}}\: \varepsilon\Bigg(\frac{k\Lambda}{\sqrt{1+(k\Lambda)^2}} \Bigg) \:,
\end{split}
\label{tau_IRS}
\end{equation}

\noindent where, $\Lambda$ is the lateral correlation length, 
$\Delta$ is the
roughness height, $N_s$ is the sheet carrier concentration,
and $N_d$ is the doping carrier density.

\subsubsection*{\bf\emph{6. \hspace{0.2cm}{Polar optical phonon  scattering}}}
\label{POP}
The POP scattering results from the interaction of optical phonons with electrons. The POP scattering mechanism is inelastic and anisotropic, which occurs via the emission or the absorption of a phonon hence, RTA is inapplicable in such SL structures. The scattering rate due to the POP scattering mechanism is approximately constant at very high energies, and it depends on the POP frequencies. The POP scattering dominates in the higher temperature domain. Hence, it is significant at both near and beyond room temperature. 
The out-scattering operator is given by \cite{rode1975low}
\begin{equation}
    S_o=\Big(N_{pop} +1 -f^-\Big)\lambda_o^- +\Big(N_{pop}+f^+\Big)\lambda_o^+ \:,
    \label{out_scattering_operator}
\end{equation}
\begin{equation}
      \lambda_o^\pm=L^\pm \Big[(A^\pm)^2 ln\Big|\frac{k^{\pm}+k}{k^{\pm}-k}\Big| -A^\pm cc^\pm -aca^\pm c^\pm\Big]\:,
      \label{lambda} 
\end{equation}
\begin{equation}
       L^\pm =\frac{e^2\: \omega_{pop}\:k^\pm }{4\pi\:\hslash\: k\:\nu(k^\pm)}\Big(\frac{\epsilon_s - \epsilon_\infty}{\epsilon_s \: \epsilon_\infty}\Big),
       \label{L} 
\end{equation}
 where, $\epsilon_\infty$ and $\epsilon_s$ are high and low-frequency dielectric constants, respectively. 
\begin{equation}
       A^\pm =aa^\pm +[(k^\pm)^2 + k^2]\: cc^\pm / \:2\: k^\pm k \:,\label{A}
\end{equation}
\noindent where c, $c^\pm$, a and  $a^\pm$ are the wave function coefficients, $k^{\pm}$ is the solution of Eq. $\varepsilon(k) \pm \hslash \omega_{pop}$. Any quantity superfixed by plus/minus is to be evaluated at the energy corresponding to $k^{+}$ or $k^{-}$. The superscript plus denotes scattering by the absorption and is evaluated at an energy $\varepsilon(k) +\hslash \omega_{pop}$. Similarly, superscript minus denotes scattering by the emission and is evaluated at energy $\varepsilon(k)-\hslash \omega_{pop}$. Emission of phonons is possible only if the phonons' energy is greater than $\hslash \omega_{pop}$ energy. Therefore, if the phonon energy is less than $\hslash \omega_{pop}$, the term $\lambda_{o}^{-}$ has to be considered as zero. The term $N_{pop}$, indicates the number of optical phonons and is given by the Bose distribution as \cite{rode1970electron,rode1975low}
\begin{equation}
    N_{pop}=\frac{1}{exp\:(\hslash\: \omega_{pop}\:/\:k_B\: T)\:-\:1} \:.\\
    \label{number_of_phonons}
\end{equation}
The in-scattering operator $S_i$, is given by
\begin{equation}
    S_i=(N_{pop}+1-f)\lambda_i^+g^+ + (N_{pop} +f)\lambda_i^{-} g^- \:,\\
    \label{in_scattering_operator}
\end{equation}
where, plus and minus superscripts indicate the absorption and emission processes, respectively. The term $\lambda_i^\pm(k)$ can be expressed as
\begin{equation}
    \begin{split}
        \lambda_i^\pm(k)=L^\pm\:\Big[\frac{(k^{\pm})^2\: +\: k^2}{2\:k^{\pm}\:k}\:(A^\pm)^2\:ln\:\Big|\frac{k^{\pm}\:+\:k}{k^{\pm}\:-\:k}\Big|\\ 
         -\:(A^\pm)^2 \:-\:\frac{ c^2(k)\:(c^\pm(k))^2}{3}\Big] \:.\\
    \end{split}
    \label{lambda_i}
\end{equation}

The mobility can be calculated after calculating the rates of all the elastic scattering mechanisms $\frac{1}{\tau_{el}(k)}$ \eqref{tau_elastic} and the influence of inelastic scattering mechanisms on $g$ \eqref{perturbation} through the terms $S_{i}(g)$ \eqref{in_scattering_operator} and $S_o$ \eqref{out_scattering_operator}. The rates of various elastic scattering mechanisms are calculated by using the expressions given in Eqs. \eqref{tau_II}, \eqref{tau_PZ}, \eqref{tau_ADP}, \eqref{tau_IRS}.

\subsection{Mobility and conductivity}
The RTA \cite{lundstrom2002fundamentals} cannot be used if the scattering process is inelastic and anisotropic because there is no way to define the relaxation time that is independent of the distribution function. In such instances, Rode's iterative approach can be applied to compute the real distribution function under low-field conditions.
After calculating the perturbation distribution by using Rode's algorithm, we finally calculate the low-field carrier mobility, $\mu$ \cite{rode1970electron,rode1975low,faghaninia2015ab}
\begin{equation}
    \mu=\:\frac{1}{3E}\frac{\int\:{\bf{\nu}}(\varepsilon)\: D_S(\varepsilon)\: g(\varepsilon)\: d\varepsilon}{\int\: D_S(\varepsilon)\:f_0(\varepsilon)\:d\varepsilon} \:.\\
    \label{mu}
\end{equation}

\indent The term $g(\varepsilon)$, can be obtained from Eq. \eqref{perturbation} and the carrier velocity $\bf{\nu}(k)$ can be calculated from the band structure as 
\begin{equation}
     \bf{\nu}(k)=\:\frac{1}{\hslash}\frac{\partial \varepsilon}{\partial \textbf{k}} \:.\\
     \label{velocity}
\end{equation}

\begin{figure}[!t]
	\centering
	{\includegraphics[height=0.55\textwidth,width=0.48\textwidth]{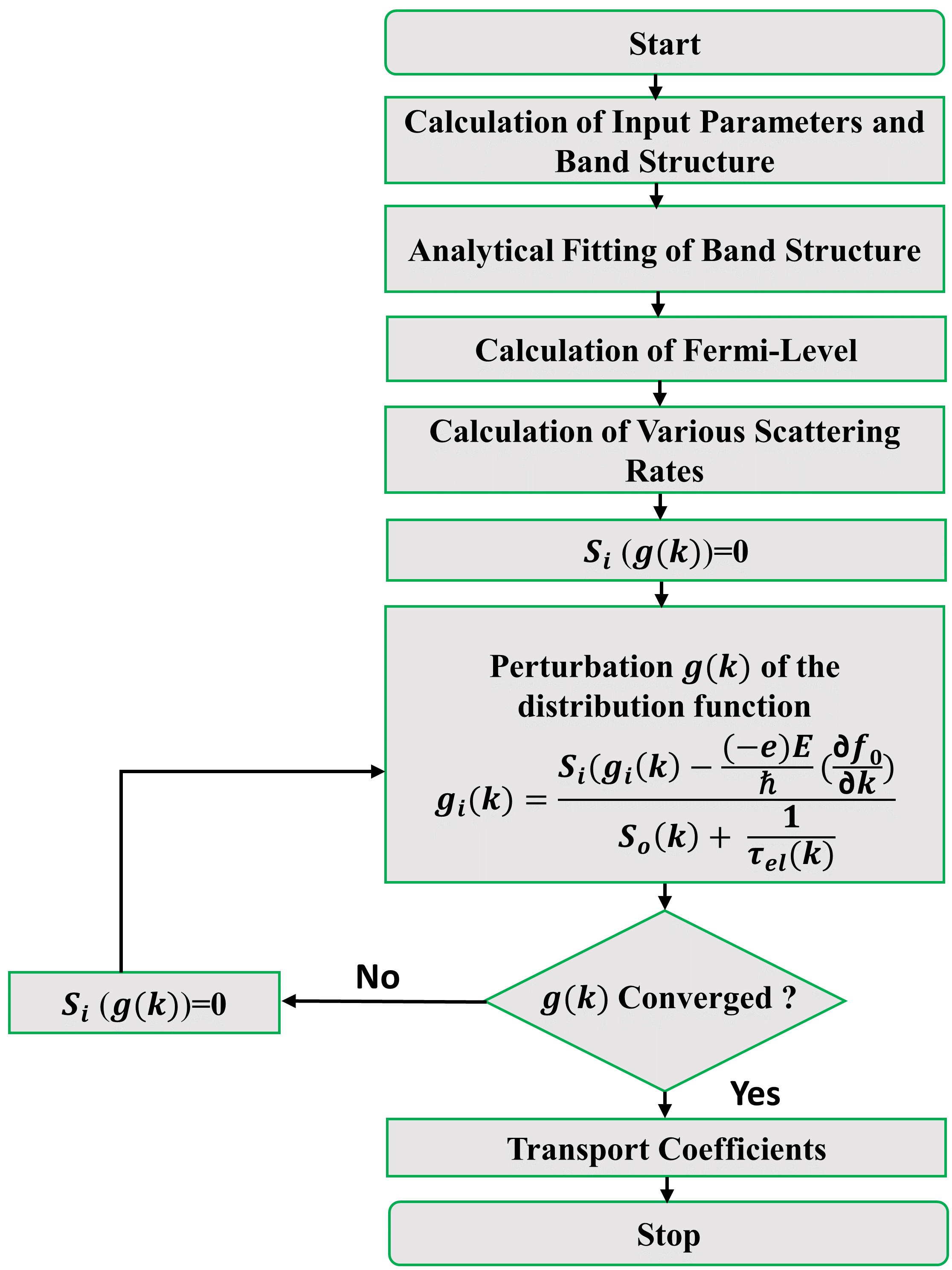}}
	\quad
	\caption{Flowchart for the calculation of electronic transport parameters.}
	\label{flowchart}
\end{figure}

Once the mobility is determined, it is pretty easy to calculate the electrical conductivity by using
\begin{equation}
     \sigma=n\:e\:\mu \:,\\
     \label{sigma}
\end{equation}

\noindent where, $\mu$ is the electron drift mobility, and $n$ is the electron carrier concentration. The entire sequence for calculating the transport coefficients using Rode's approach is shown in Fig. \ref{flowchart}.

Similarly, in the presence of an arbitrary magnetic field, the BTE can be solved. The distribution function in such cases can be written as \cite{rode1973theory,mandia2022electrical} 
\begin{equation}
    f(\textbf{k})=f_0[\varepsilon(k)]+xg(k)+yh(k)\:,\\
    \label{distribution_func_in_magneticfield}
\end{equation} 

\noindent where, $y$ is the direction, cosine from $\textbf{B}\times \textbf{E}$ to $\textbf{k}$, and $h(k)$ is the perturbation distribution function due to the magnetic field. Substituting Eq. \eqref{distribution_func_in_magneticfield} in \eqref{BTE} gives a pair of coupled equations that can be solved iteratively \cite{rode1973theory}

\begin{equation}
    g_{i+1}(k)=\frac{S_i(g_i(k) - \frac {(-e)E}{\hslash}\:(\frac{\partial f_0}{\partial k}) + \beta S_i(h_i(k))}{S_o(k)\:(1+\beta^2)} \:,\\
    \label{perturbation due to E field}
\end{equation} 
\begin{equation}
    h_{i+1}(k)=\frac{S_i(h_i(k) + \beta\:\frac {(-e)E}{\hslash}\:(\frac{\partial f_0}{\partial k}) - \beta S_i(g_i(k))}{S_o(k)\:(1+\beta^2)} \:,\\
    \label{perturbation due to magnetic field}
\end{equation} 

\noindent where, $\beta=\frac {(-e)\nu (k) B}{\hslash k S_o(k)}$, and $B$ is the applied magnetic field. The expression for the Hall mobility and the Hall scattering factor can be written as \cite{vasileska2017computational}
\begin{equation}
    \mu_{_H}=\:\frac{1}{B}\frac{\int\:{\bf{\nu}}(\varepsilon)\: D_S(\varepsilon)\: h(\varepsilon)\: d\varepsilon}{\int\:{\bf{\nu}}(\varepsilon)\: D_S(\varepsilon)\:g(\varepsilon)\:d\varepsilon} \:,\\
    \label{Hall_mobility}
\end{equation}

\begin{equation}
     r_{_H}=\:\frac{\mu_{_H}}{\mu} \:,\\
     \label{Hallscattering_factor}
\end{equation}

\noindent where, $\mu_H$ and $\mu$ are the Hall and the drift mobility, respectively, and $r_{_H}$ is the Hall scattering factor. This solution gives a more accurate result for the Hall scattering factor compared with the other expressions based on the RTA \cite{mandia2022electrical}.

\section{Simulation approach}
\label{simulation_approach}


First, we calculate the band structure using the {\bf k.p} technique as discussed in Sec. \ref{k.p} and then analytically fit it
to produce a smooth curve for the calculation of group velocity \cite{mandia2021ammcr}. By using Eq. \eqref{fermi_level}, the Fermi level is determined with a smooth band structure obtained after the analytical fitting, where $V_0$ represents the volume of the cell and $\varepsilon_c$ represents the energy at the bottom of the CB.
\begin{equation}
     n=\frac{1}{V_0}\int_{\varepsilon_c}^{\infty}D_S(\varepsilon)f(\varepsilon) d\varepsilon\:.\\
     \label{fermi_level}
\end{equation}

 Equations \eqref{tau_II}, \eqref{tau_PZ}, \eqref{tau_ADP}, \eqref{tau_IRS}, \eqref{out_scattering_operator}, \eqref{in_scattering_operator} are used to calculate the various scattering rates, and the perturbation in the distribution function is determined using Eq. \eqref{perturbation} with $S_i(k)=0$.
 The term $g(k)$, is calculated iteratively until $g(k)$ 
 converges and it gives results beyond the RTA. 

\section{Results and Discussion}
\label{sec_result}
\subsection{Dispersion relation for T2SL}
\begin{figure}[!htbp]
	\centering
	\subfigure[\:(110)]{\includegraphics[height=0.231\textwidth,width=0.225\textwidth]{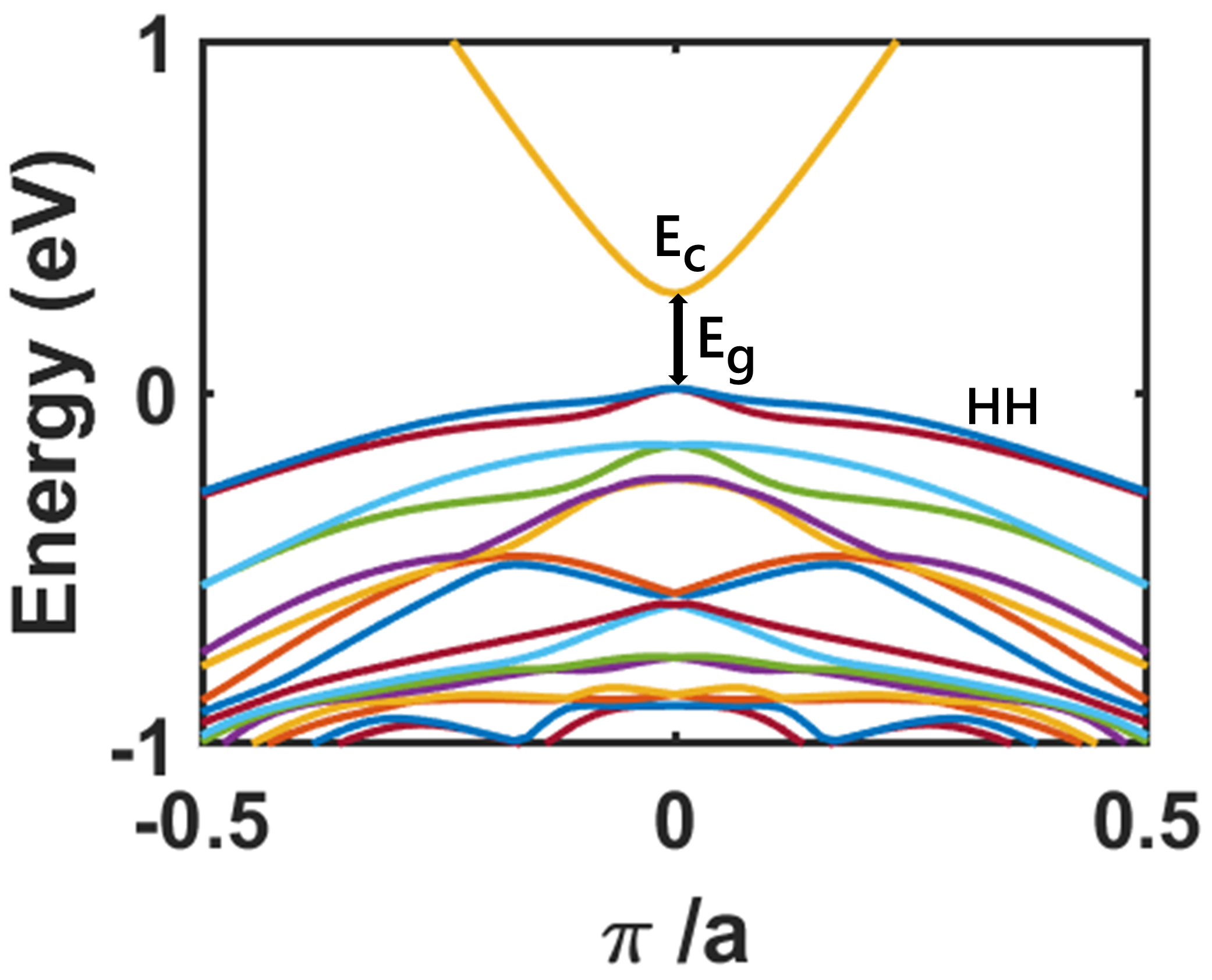}\label{Eka}}
	\quad
	\subfigure[\:(001)]{\includegraphics[height=0.225\textwidth,width=0.225\textwidth]{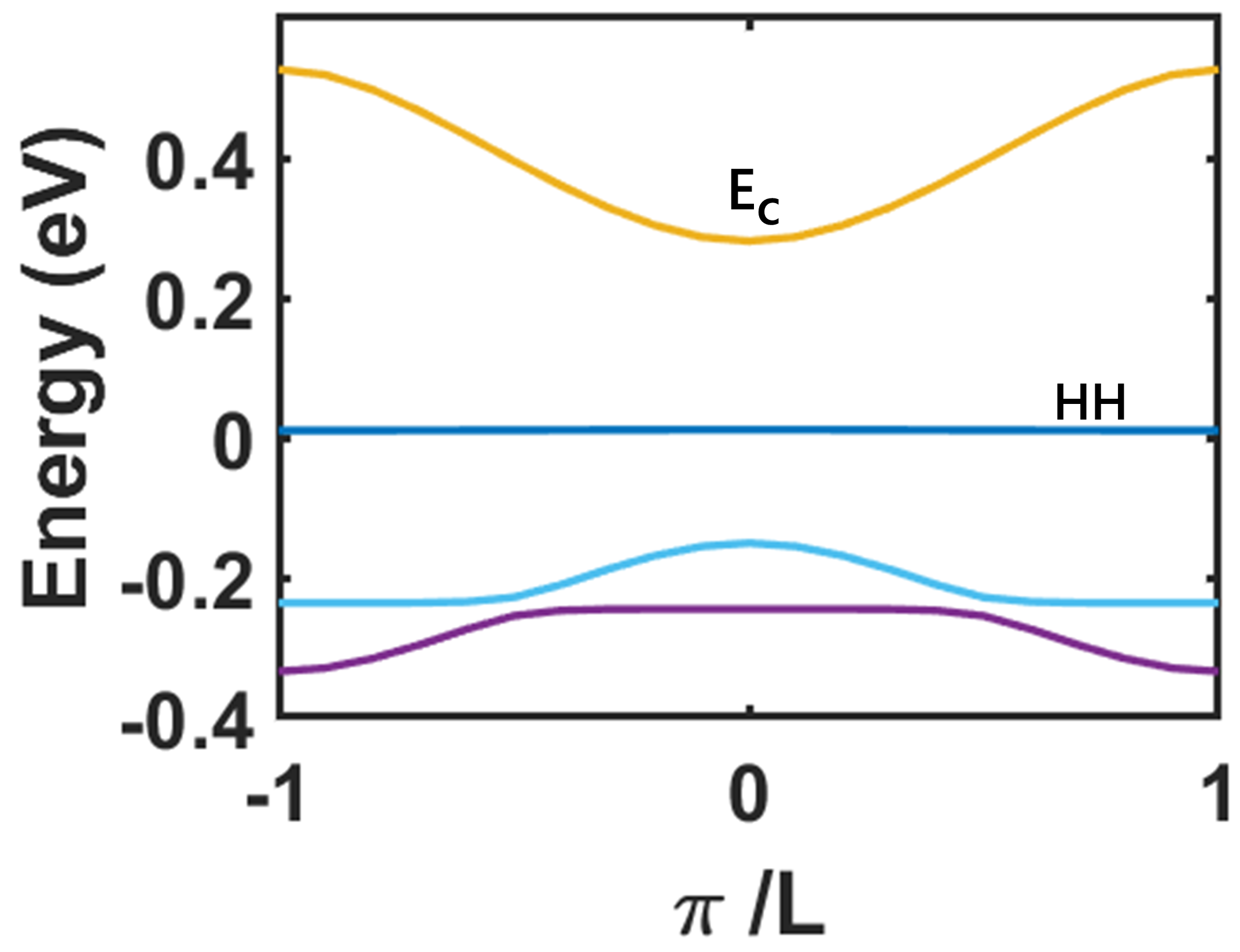}\label{Ekb}}
	\quad
	\caption{Calculated band structure in the first BZ using the periodic boundary condition of a T2SL based on 8 ML InAs / 8 ML GaSb at $T=77$ K using the $\bf{k.p}$ method (a) The in-plane dispersion and (b) the out-of-plane dispersion.}
	\label{bandstructure}
\end{figure}

\begin{table*}[t]
\caption{\label{table1}
Material parameters required to calculate the electronic band structure using the $\bf{k.p}$ technique at T = 77 K \cite{livneh2012k,becer2019modeling,vurgaftman2001band,delmas2019comprehensive} }
\begin{ruledtabular}
\begin{tabular}{cccc}
\bf{Quantity} & \bf{Unit} & \bf{InAs} & \bf{GaSb}  \\ \hline

Lattice constant & {\AA} & 6.0584& 6.0959 \\

Effective mass of electron ($m_e^*$) & -& 0.022&0.0412\\

Energy band gap at 0 K & $eV$ & 0.418 & 0.814 \\
Luttinger parameter $\gamma1$ & - & 19.4& 11.84 \\
Luttinger parameter $\gamma2$ & - & 8.545 & 4.25\\
Luttinger parameter $\gamma3$  & -& 9.17 & 5.01\\
Varshini Parameter $\alpha $ & $meV/K$ & 0.276 & 0.417\\
Varshini Parameter $\beta$ & $K$ & 93 & 140\\
Interband mixing parameter Ep & $eV$ & 21.5 & 22.4\\
Spin-orbit splitting (SO) & $eV$ & 0.38& 0.76\\
Valence band offset (VBO) & $eV$ & -0.56 & 0\\

\end{tabular}
\end{ruledtabular}
\end{table*}

\begin{figure}[!htbp]
	\centering
	{\includegraphics[height=0.32\textwidth,width=0.43\textwidth]{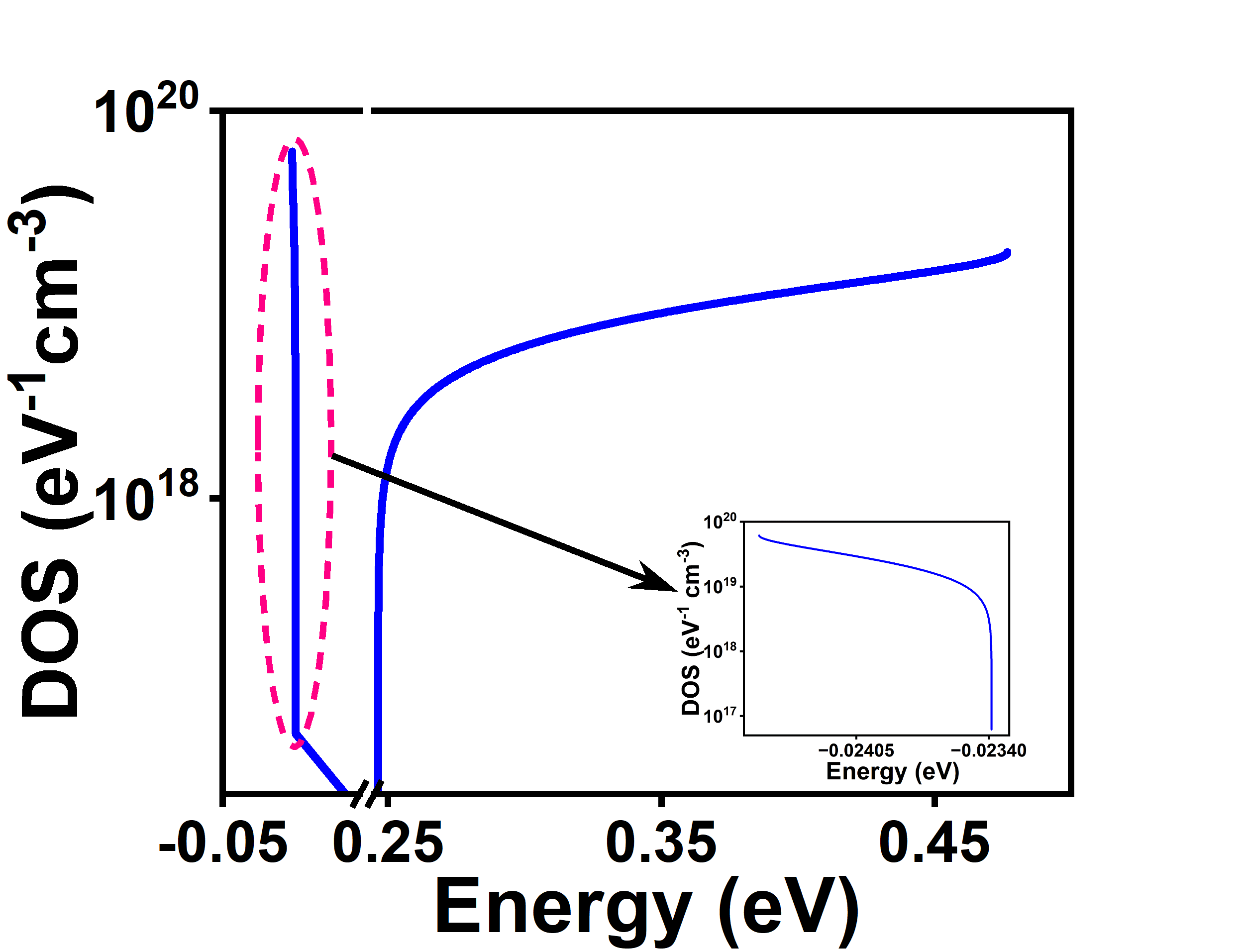}}
	\quad
	\caption{DOS calculated using the $\bf{k.p}$ method in an InAs/GaSb SL as a function of energy. The inset clearly shows how the DOS for the carriers in the VB varies as a function of energy.}
	\label{DOS}
\end{figure}

\begin{figure*}[!htbp]
	\centering
	\subfigure[\hspace{0.1cm} {T=77 K}]{\includegraphics[height=0.25\textwidth,width=0.30\textwidth]{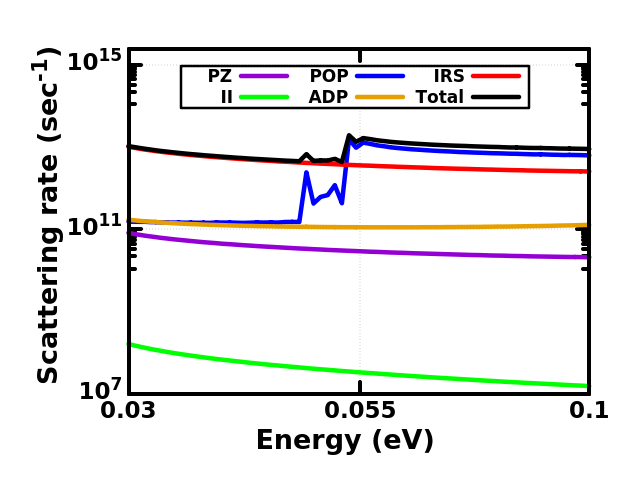}\label{77K_a}}
	\quad
	\subfigure[\hspace{0.1cm} {T=300 K}]{\includegraphics[height=0.25\textwidth,width=0.30\textwidth]{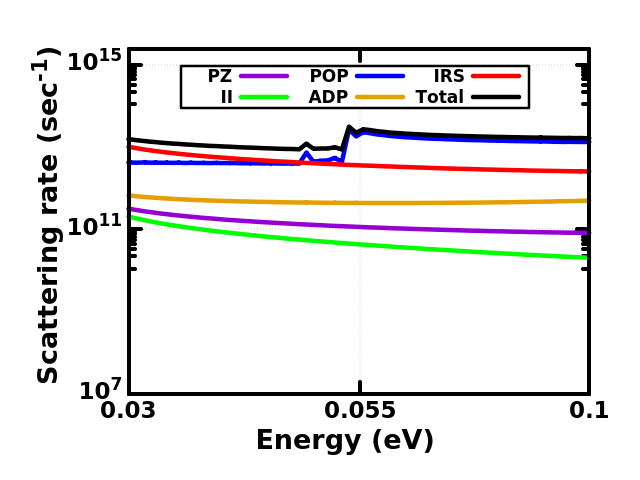}\label{300K_a}}
	\quad
	\subfigure[\hspace{0.1cm} {T=500 K}]{\includegraphics[height=0.25\textwidth,width=0.30\textwidth]{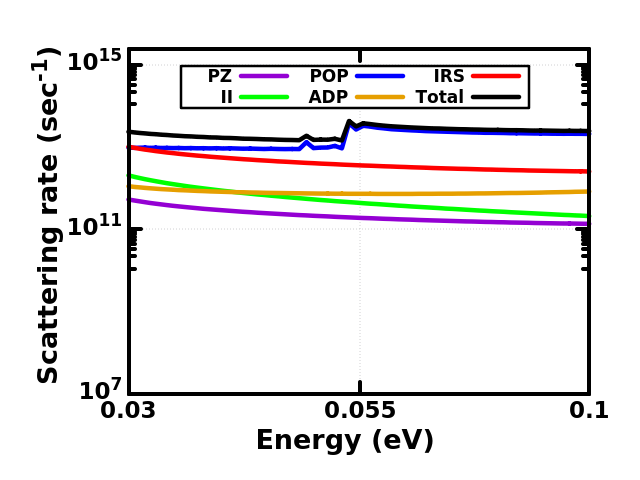}\label{500K_a}}
	\quad
	\subfigure[\hspace{0.1cm} {T=77 K}]{\includegraphics[height=0.25\textwidth,width=0.30\textwidth]{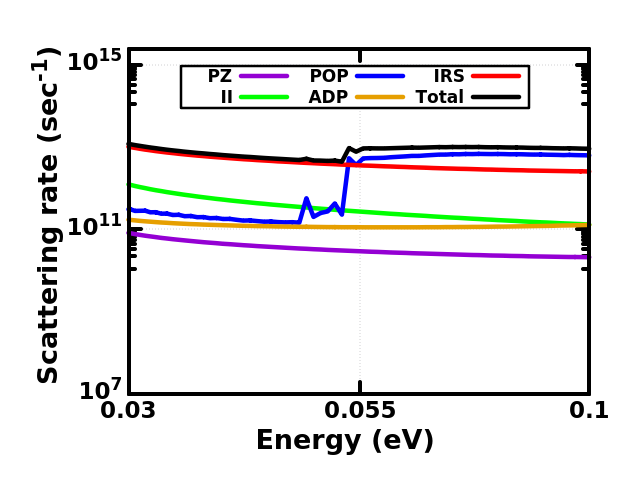}\label{77K_b}}
	\quad
	\subfigure[\hspace{0.1cm} {T=300 K}]{\includegraphics[height=0.25\textwidth,width=0.30\textwidth]{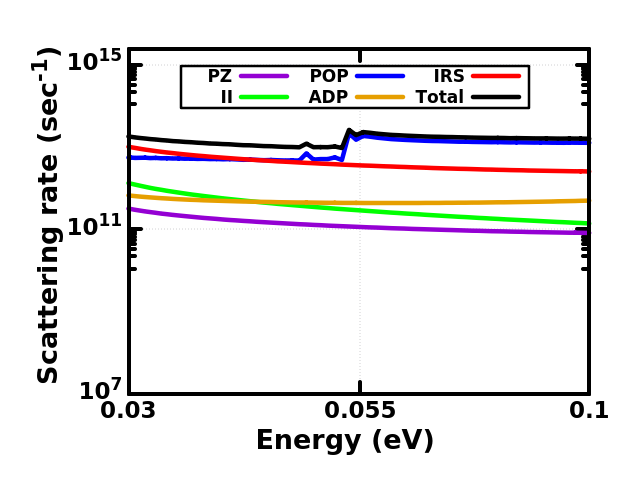}\label{300K_b}}
	\quad
	\subfigure[\hspace{0.1cm} {T=500 K}]{\includegraphics[height=0.25\textwidth,width=0.30\textwidth]{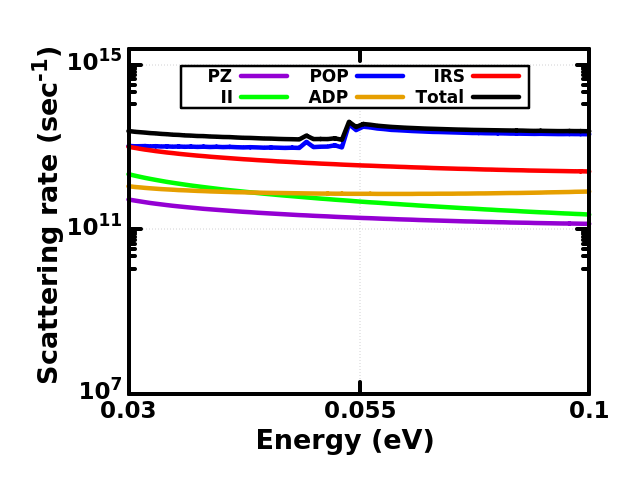}\label{500K_b}}
\quad
\caption{Scattering rates for 8ML/8ML InAs/GaSb based T2SL with roughness parameters ${\Lambda=3 \:nm}$ and ${\Delta=0.3 \:nm}$ as a function of electron energy at  
  \hspace{0.05cm} (a) $T=77\:K\: and \:N_{D}=1\times 10^{13}      \:cm^{-3}$\hspace{0.1cm} (b) $T=300\:K\: and\: N_{D}=1\times 10^{13}   \:cm^{-3}$\hspace{0.1cm} (c) $T=500\:K\: and\: N_{D}=1\times 10^{13}   \:cm^{-3}$\hspace{0.1cm} (d) $T=77\:K\: and\: N_{D}=2\times 10^{17}   \:cm^{-3}$\hspace{0.1cm} (e)  $T=300\:K \:and\: N_{D}=2\times 10^{17}   \:cm^{-3}$\hspace{0.1cm} and (f) \hspace{0.05cm}  $T=500\:K\: and\: N_{D}=2\times 10^{17}   \:cm^{-3}$\hspace{0.1cm}.}
\label{scattering_rate}
\end{figure*}

We calculate the band structure of an InAs/GaSb-based T2SL, with layer widths nML/mML, where n, m = 8, 8 correspondingly, using the $8\times8$ {\bf k.p} technique as described in Sec. \ref{k.p}, at a temperature of T=77 K, and the results are shown in Fig. \ref{bandstructure}. In a single period of 8ML/8ML InAs/GaSb configuration, the thickness of each layer is roughly 24 \AA. The dispersion curve along the in-plane and the out-of-plane directions are presented in Figs. \ref{Eka} and \ref{Ekb}, respectively and the calculated band gap is 270 meV. The band gap of 270 meV corresponds to a cut-off wavelength of 4.59 $\mu$m which confirms that our model is best suited for the MWIR spectrum. In Fig. \ref{DOS} we show the DOS of an SL as a function of energy, calculated using the {\bf k.p} method. Table \ref{table1} summarizes the values of the parameters, utilized in the {\bf k.p} calculations.

\subsection{Scattering rates}




\begin{figure}[!htbp]
	\centering
	{\includegraphics[height=0.32\textwidth,width=0.44\textwidth]{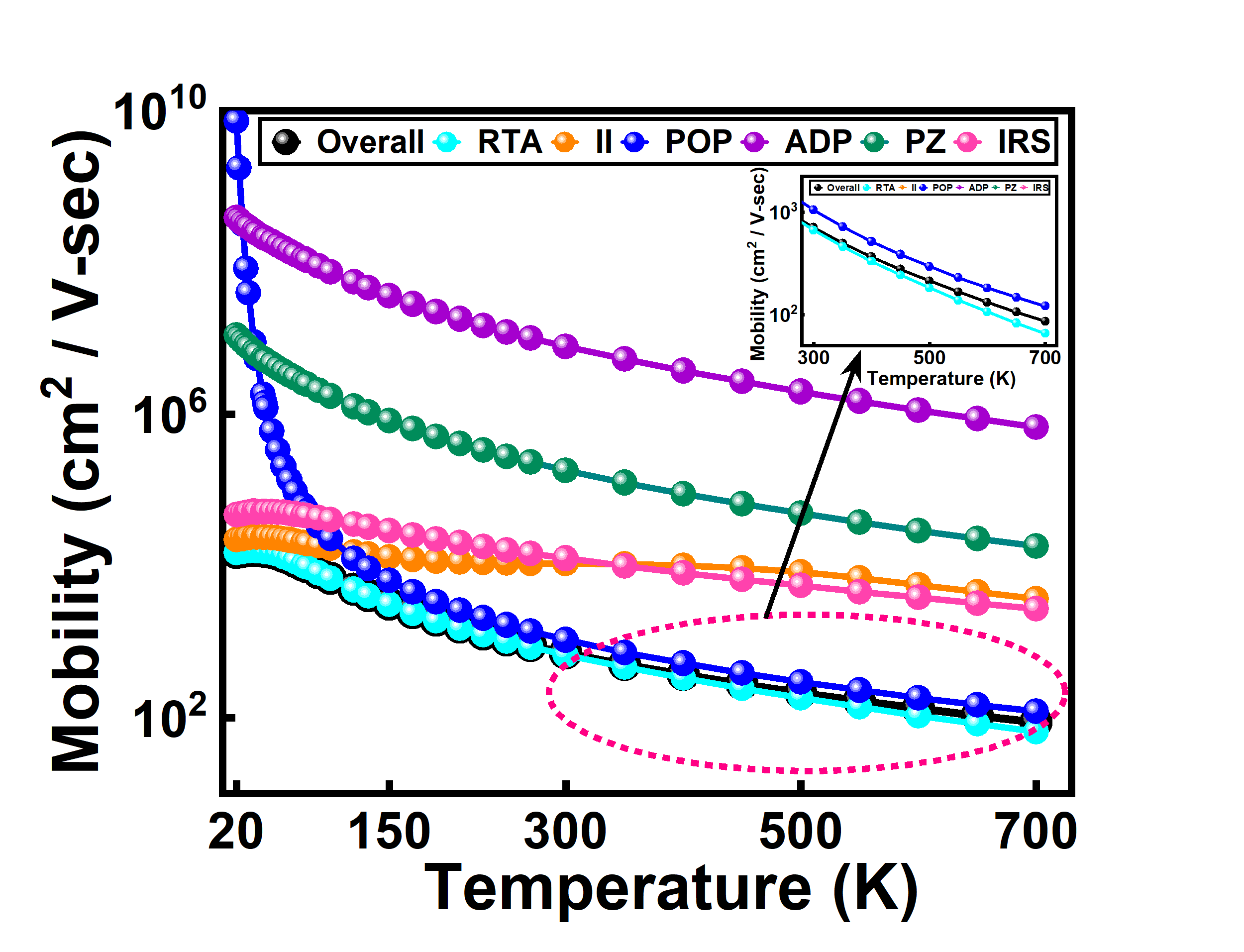}}
	\quad
    \caption{Calculated mobility contribution for electrons due to the various scattering mechanism involved in (8ML/8ML) InAs/GaSb T2SL as a function of temperature for $N_{D}=9\times10^{16}~cm^{-3}$.}
\label{mob_contribution}
\end{figure}

\begin{figure}[!htbp]
	\centering
	{\includegraphics[height=0.3\textwidth,width=0.4\textwidth]{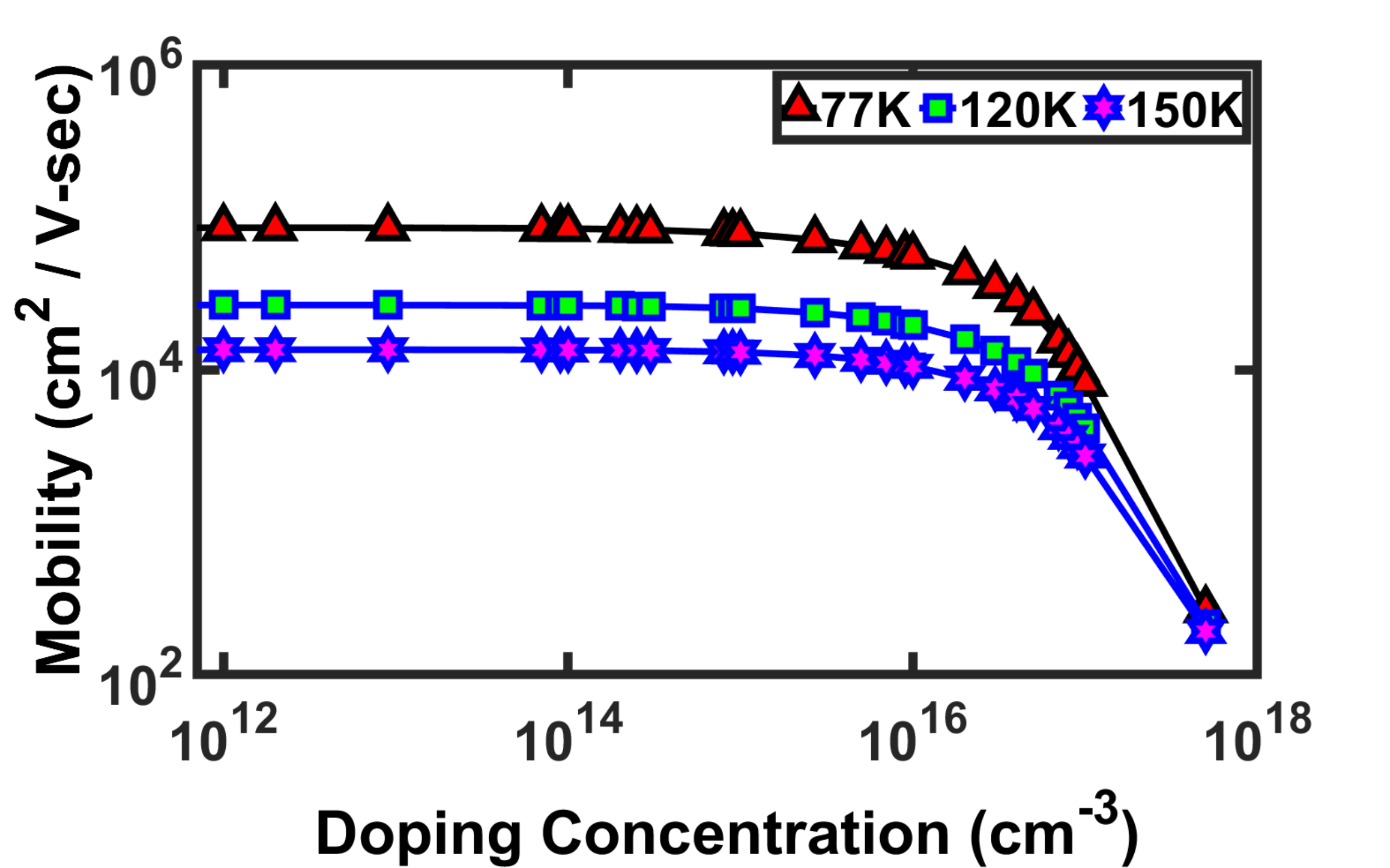}}
	\quad
   	\caption{Calculated low-field electron drift mobility in 8ML/8ML InAs/GaSb SL as a function of doping concentration for temperatures of 77 K, 120 K and 150 K.}
\label{Mob_dop}
\end{figure}

\begin{table*}
\caption{\label{tab:Parameters}Material parameters required to compute the various scattering rates \cite{vurgaftman2001band,rode1970electron,mitra1963phonon,lockwood2005optical,haugan2011design,alchaar2019characterization}.}
\begin{ruledtabular}
\begin{tabular}{cccc}
 \textbf{Parameter} & \textbf{Unit} & \textbf{InAs} & \textbf{GaSb}\\ \hline

 Elastic constant $c_{11}$ & GPa & 832.9 & 884.2 \\
 Elastic constant $c_{12}$ & GPa & 452.6 & 402.6 \\
 Elastic constant $c_{44}$ & GPa & 395.9 & 432.2 \\
 Acoustic deformation potential & eV & 4.90 & 6.70 \\
 Low freq. dielectric constant & - & 14.55 & 15.00 \\
 High freq. dielectric constant & - & 11.78 & 13.80 \\
 Piezoelectric coefficient & $C/m^2$ & 0.045 & 0.126 \\
 Optical phonon frequency & 1/cm & 240 (LO)\footnote{LO~: Longitudinal\:Optical\: Phonon\: Frequency.}, 218 (TO)\footnote{TO~: Transverse\: Optical\: Phonon\: Frequency.} & 193 (LO)\footnotemark[1], 215 (TO)\footnotemark[2] \\
 
\end{tabular}
\end{ruledtabular}
\end{table*}

\indent In Fig. \ref{scattering_rate}, we show the dependence of scattering rates with energy for the temperatures of 77 K, 300 K, and 500 K at doping densities of $N_D =1\times 10^{13}~cm^{-3}$ and $N_D =2\times 10^{17}~cm^{-3}$. Here, we show the relative importance of each of the scattering mechanisms in a T2SL. The IRS mechanism is the strongest scattering mechanism for low as well as high doping densities at a temperature of 77 K and 300 K as shown in Fig. \ref{scattering_rate}. At a temperature of 77 K and a doping density of $N_D =1\times 10^{13} \: cm^{-3}$, the most dominant contributions are due to the IRS followed by the ADP and the POP scattering mechanisms. The II scattering mechanism is the least significant scattering mechanism at this particular temperature and doping density, whereas it has a significant contribution at higher doping densities.

\begin{figure}[t]
	\centering
	{\includegraphics[height=0.32\textwidth,width=0.44\textwidth]{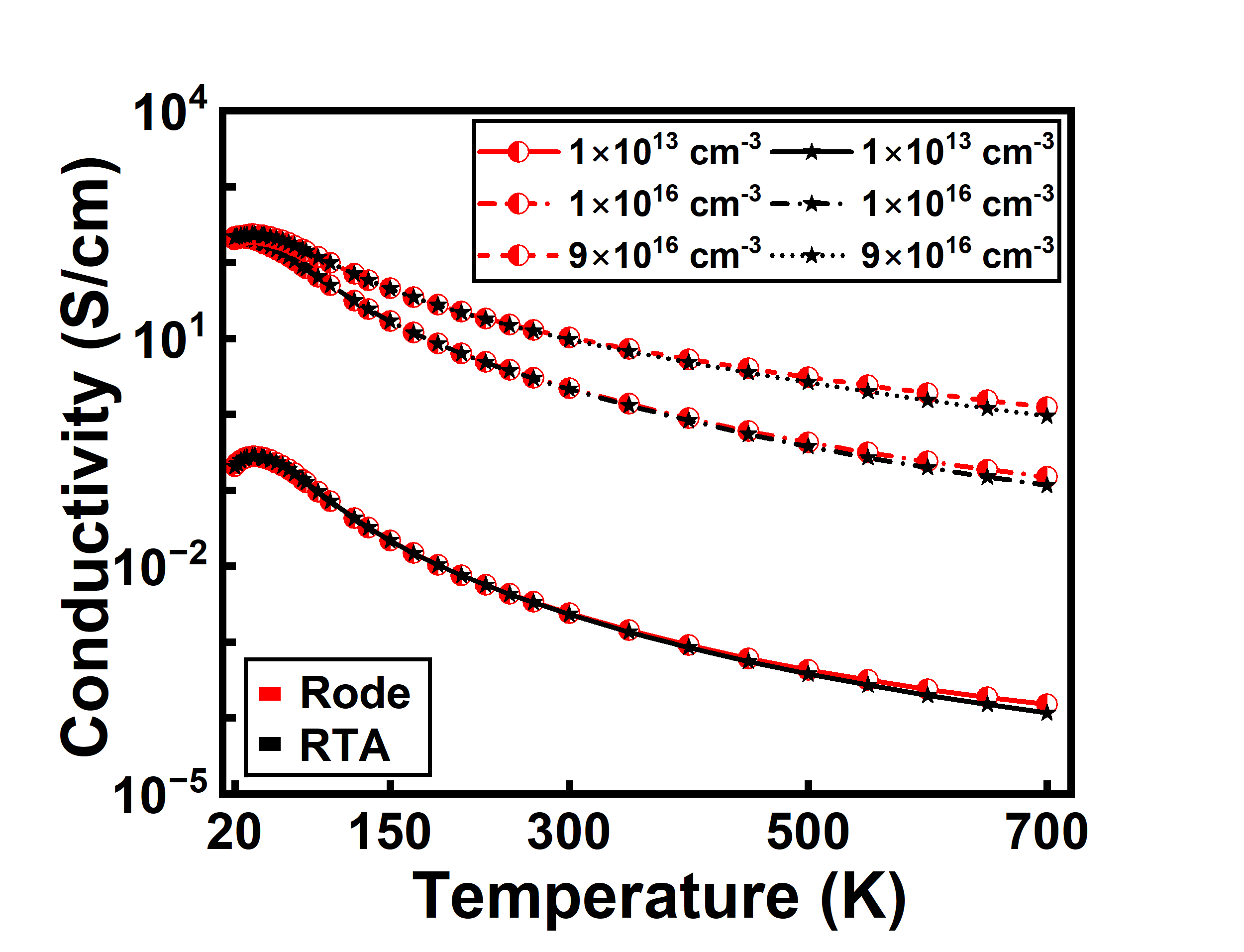}}
	\quad
	\caption{Comparison of conductivity in a T2SL as a function of temperature, calculated using the Rode's and the RTA method for various doping concentrations.}
	\label{conductivity}
\end{figure}

\begin{figure}[!t]
	\centering
	{\includegraphics[height=0.32\textwidth,width=0.43\textwidth]{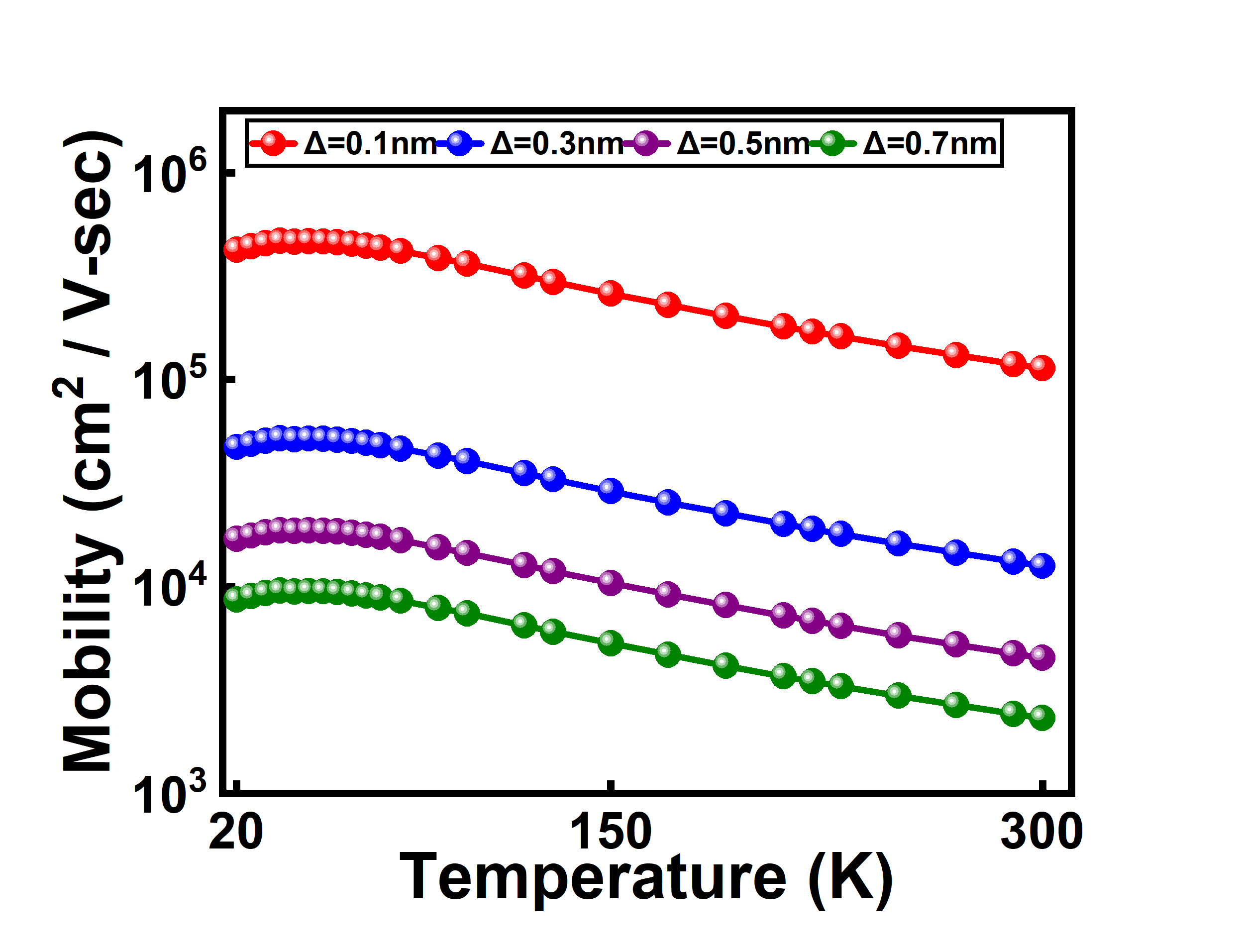}}
	\quad
	\caption{Calculated temperature dependence of electronic mobility with IRS heights for a correlation length of 3 nm \& $N_{D}=9\times10^{16}~{cm^{-3}}$. The mobility due to only the IRS mechanism is shown.}
	\label{MOb_IRS height}
\end{figure}

\indent At room temperature, the average energy of the carriers is $3/2 k_B T= 0.0388~eV$, indicating that the majority of the carriers are in the low-energy region. Hence, it is clear from Fig. \ref{300K_b} that at room temperature, the significant contribution comes from the IRS mechanism as well as the POP scattering mechanism. Both scattering mechanisms are dominant at this temperature, and the dominance of the POP scattering mechanism changes with respect to temperature and the average energy of the carriers, which signifies that the POP scattering mechanism plays a significant role in such a T2SL structure. As a result, it is important to note that the POP scattering mechanism is the primary factor limiting the carrier's mobility from room temperature to higher temperatures.\\
\indent At a temperature of 500 K, the average energy of the carriers is $0.0646~eV$ and, most of the carrier contributes to the POP scattering mechanism hence, this again demonstrates that the POP scattering mechanism is the most dominant scattering mechanism for T2SL at and beyond the ambient temperature for both doping densities, as shown in Figs. \ref{500K_a} and \ref{500K_b}. Figure \ref{scattering_rate} shows a sudden change in the POP scattering rate after particular energy, which is because if the electron energy is less than the POP energy, the electron can only scatter by the absorption of the optical phonons, whereas if the energy is greater than the phonon energy, the electron can scatter by both the absorption and the emission of phonons, where the optical phonon energy is determined using $\hslash \omega_{_{POP}}$. The PZ scattering is the least dominant scattering mechanism at higher doping densities, as shown in Figs. \ref{77K_b}, \ref{300K_b}, \ref{500K_b}. Table \ref{tab:Parameters}  lists the material parameters that are used to compute the various scattering rates.\\
\indent It is generally known that the ADP scattering mechanism becomes substantial at temperatures of 77 K and above, reducing electron mobility. Therefore, it is also important to include the effect of the ADP scattering mechanism, which is significant near the room temperature for low as well as high doping densities, which was not highlighted in the earlier works for such SL structures. At lower temperatures and in the thin-film systems, the IRS scattering is considerable, and to compute the roughness scattering rate, we utilize a sheet carrier density $N_s$, of $4.6 \times 10^{12}~cm^{-2}$ and a doping carrier density $N_d$, of $1\times 10^{11}~cm^{-2}$ with the roughness height $\Delta$, fixed at $0.3~nm$, and the correlation length of the fluctuations $\Lambda$ kept at $3~nm$. The IRS mechanism is temperature independent, but the carrier distribution function depends on the temperature. Therefore, the electron mobility through the IRS mechanism is somewhat temperature sensitive. Except for the IRS scattering rate, which is temperature independent, we see that all the scattering rates increase as the temperature rises as shown in Figs. \ref{77K_a}, \ref{300K_a}, \ref{500K_a}. When the temperature is either low or intermediate, the II scattering rate increases with an increase in the doping concentration, which suppress the contribution from the PZ scattering, as shown in Figs. \ref{77K_a}, \ref{77K_b}, \ref{300K_a}, \ref{300K_b}.
\subsection{Electron transport parameters}
\indent We calculate the mobility and the conductivity for a T2SL at various temperatures and doping concentrations. Figure \ref{mob_contribution} shows the contribution to the mobility due to various scattering mechanisms calculated for $N_D=9\times10^{16}\:cm^{-3}$. To the best of our knowledge, the combined effect of these scattering mechanisms in a T2SL structure has never been shown in earlier works. These five types of scattering mechanisms show their significant contribution to the overall mobility calculation. From Fig. \ref{mob_contribution} it turns out that the scattering mechanism with the lowest mobility values is the dominant one in that temperature range. Therefore, starting at a temperature of 150 K, the POP scattering mechanism is the most dominant scattering mechanism until 700 K; below 77 K, a significant contribution to the mobility comes from the II scattering and the IRS mechanisms as shown in Fig. \ref{mob_contribution}.

\begin{figure}[!t]
    \centering
    {\includegraphics[height=0.35\textwidth,width=0.44\textwidth]{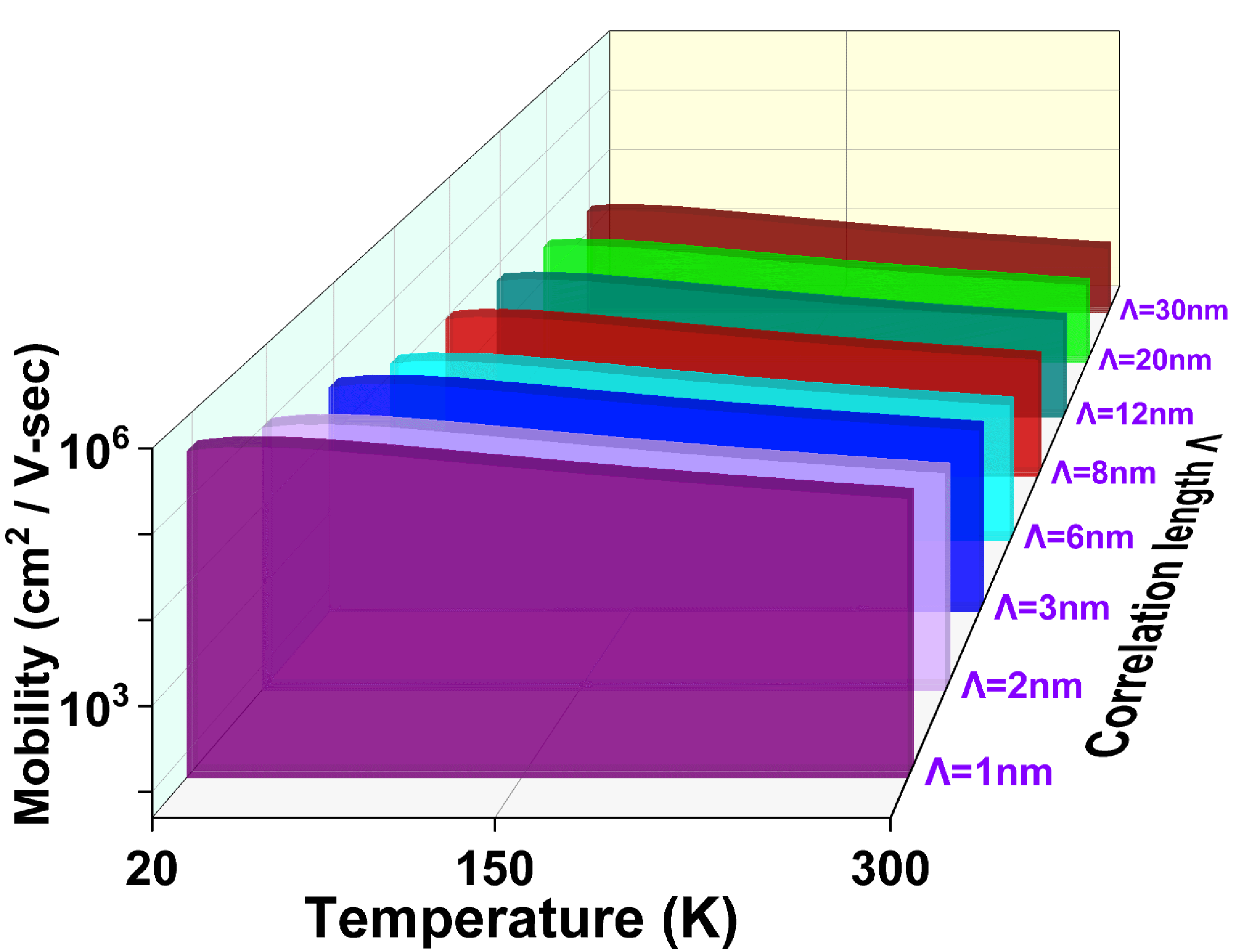}}
    \quad
    \caption{Calculated mobility for electrons in an 8ML InAs/8ML GaSb SL as a function of temperature and correlation length for an IRS height of 0.3 nm with $N_{D}=9\times10^{16}~{cm^{-3}}$. Here, the mobility due to only the IRS mechanism is shown.}
    \label{MOb_corr_length}
\end{figure}

\begin{figure}[t]
	\centering
	{\includegraphics[height=0.32\textwidth,width=0.43\textwidth]{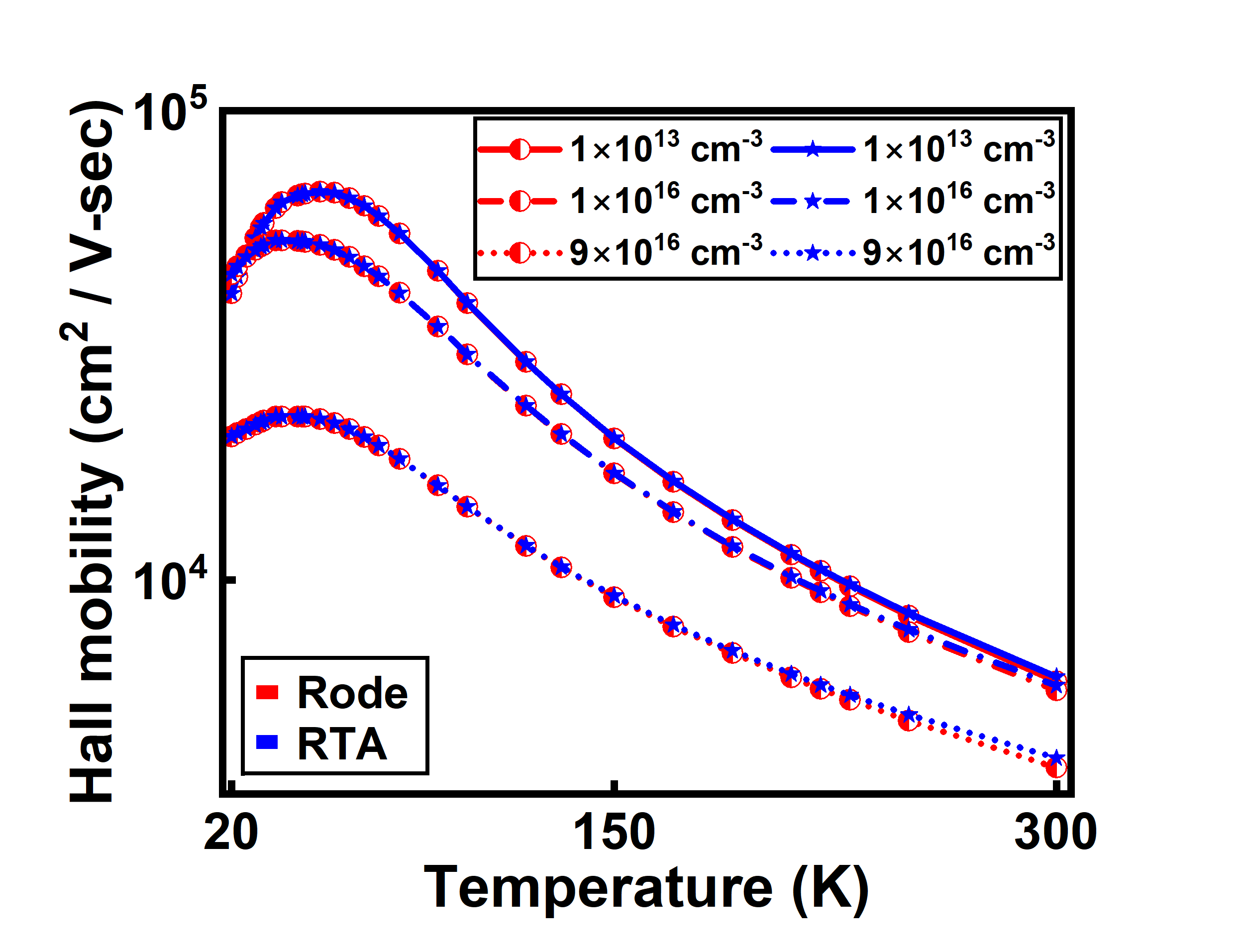}}
	\quad
	\caption{Temperature dependence of electron Hall mobility in a T2SL calculated using the Rode's and the RTA method at $B= 0.69~T$ for various doping concentrations.}
	\label{Hall_mob}
\end{figure}

\indent In case of II scattering mechanism, with increasing temperature, the electron density increases exponentially and causes growth in the screening length. As a result, the mobility at low temperatures increases sharply with rising temperatures because the scattering rates are inversely related to the square of the screening length. Since the POP scattering mechanism is more prominent above 150 K; hence the overall mobility is reduced as shown in Fig. \ref{mob_contribution}. In Fig. \ref{mob_contribution}, we also compare the mobility computed using the RTA approach to the overall mobility calculated using Rode's method and it is found that in the RTA approach, the mobility is underestimated because the POP scattering mechanism is inelastic and nonrandomizing, making it impossible to characterize the perturbation in the distribution function using the relaxation time. The POP scattering mechanism becomes insignificant at low temperatures, resulting in nearly comparable mobilities determined using the RTA and Rode's iterative technique.

\begin{figure}
	\centering
	{\includegraphics[height=0.32\textwidth,width=0.44\textwidth]{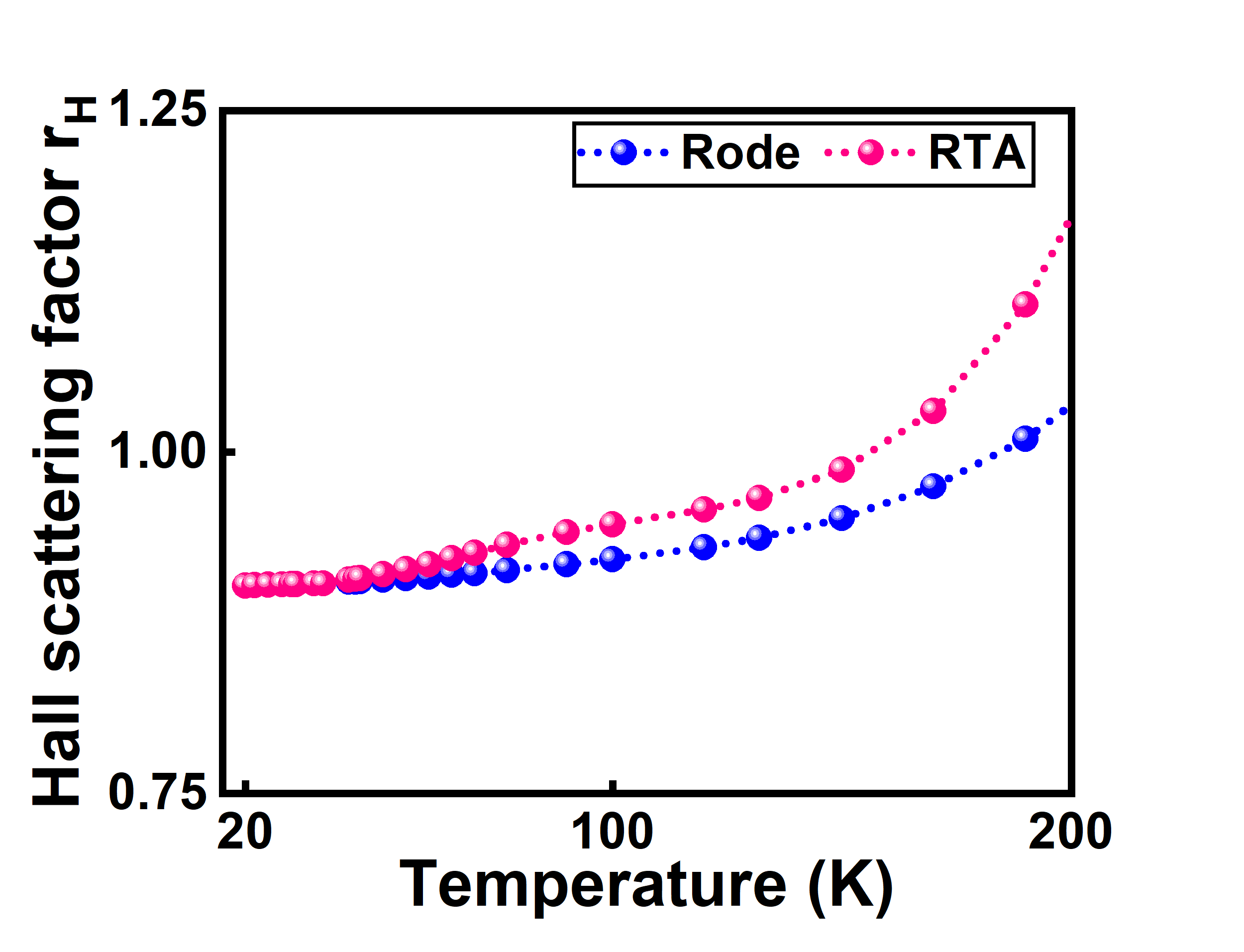}}
	\quad
	\caption{Hall scattering factor versus temperature at B= 0.69 T for $N_{D}=9\times10^{17} {cm^{-3}}$.}
	\label{Hall_factor}
\end{figure}

\begin{figure}[t]
	\centering
	{\includegraphics[height=0.32\textwidth,width=0.43\textwidth]{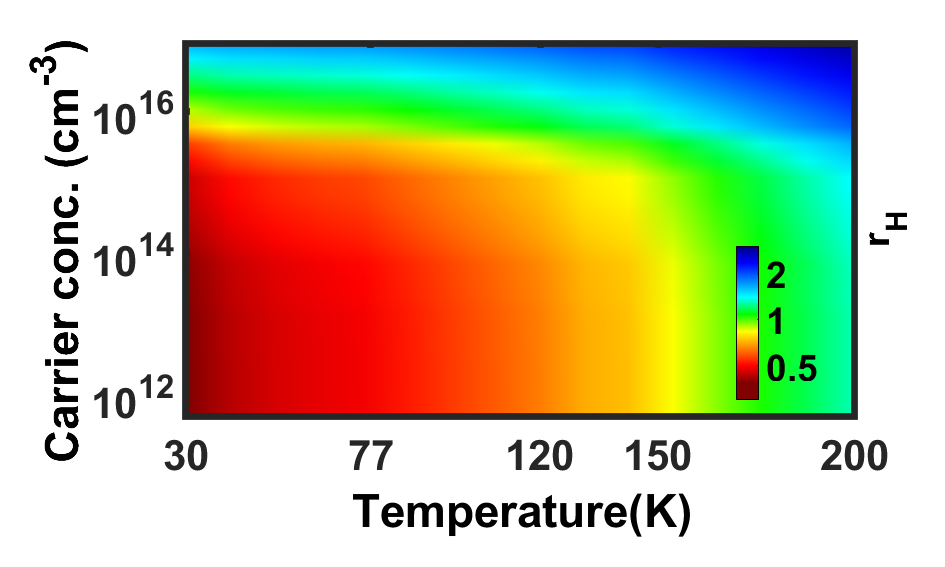}}
	\quad
	\caption{Hall scattering factor as a function of temperature and carrier concentration at B= 0.69 T.}
	\label{Hall_factor2}
\end{figure}

In Fig. \ref{Mob_dop}, we demonstrate the overall mobility versus doping concentration at different temperatures and emphasize on the mobility at $77\:K$, which is the usual operating temperature of most high-performance IR detectors. The graph illustrates a decrease in mobility as the doping concentration increases due to a rise in the number of ionized centers. As we raise the temperature, the mobility diminishes as expected because at higher temperatures the phonon scattering increases. The mobility values do not differ significantly for low carrier concentrations because the II scattering mechanism is less significant at this range and the primary contributions for lower doping concentration at low temperatures come from the PZ and the ADP scattering mechanisms, while at greater doping concentrations, the II scattering mechanism is comparable to the ADP and the PZ scattering mechanisms. The mobility owing to the II scattering mechanism is a decreasing function of $N_D$, the mobility begins to decrease as $N_D$ exceeds $1\times10^{16}\:cm^{-3}$.


\indent In Fig. \ref{conductivity}, we show the conductivity versus temperature for the doping concentrations of  $N_D=1\times10^{13}\:cm^{-3}$, $N_D=1\times10^{16}\:cm^{-3}$ and $N_D=9\times10^{16}\:cm^{-3}$, respectively, and to demonstrate the supremacy of our approach, we compare the results obtained using both the Rode's and the RTA method. At higher temperatures, the difference in the result of Rode's method and the RTA is due to the POP scattering mechanism, the POP scattering is weaker at lower temperatures hence both the RTA and the Rode exhibit the same conductivity. We demonstrate that the conductivity in a T2SL increases with an increase in the carrier concentration but decreases as we increase the temperature.

In Figs. \ref{MOb_IRS height} and \ref{MOb_corr_length}, we show the mobility due to only the IRS mechanism. The calculated mobilities are vital functions of the roughness parameters and the carrier scattering. The existing mobility calculations reveal that, up to temperatures where the POP scattering mechanism takes over, the IRS is the dominating scattering mechanism in T2SL. The screening is included in our calculation using Thomas-Fermi screening which lowers the scattering rates and increases the mobility. As illustrated in Fig. \ref{MOb_IRS height}, the mobility is shown to be strongly reliant on the roughness height $\Delta$, and decreases monotonically with increasing $\Delta$, and is proportional to $\Delta^{-2}$.

Figures \ref{MOb_IRS height} and \ref{MOb_corr_length} show that at low temperatures, the mobility rises since the value of $\frac{\partial f}{\partial \varepsilon}$ is an ascending function of temperature and the denominator of Eq. \eqref{mu} is virtually constant at lower temperatures. Also, the electron density increases at higher temperatures and hence the mobility drop smoothly. Figure \ref{MOb_corr_length} shows that the mobility is high for smaller values of correlation length $\Lambda$, and drops rapidly as the correlation length of roughness increases until it reaches a saturation point. The mobility reaches its maximum value at roughly 50 K for smaller values of $\Lambda$, and this maximum point moves toward the higher temperatures for greater values of $\Lambda$.


\indent The Hall mobility in InAs/GaSb T2SLs is depicted in Fig. \ref{Hall_mob}. At temperatures above 50 K, the mobility reduces as expected from a combination of the ADP and the POP scattering mechanisms. In T2SL, the mobility increases with decreasing temperature, preferable to the $T^{-3/2}$ dependency associated with the phonon scattering. The greater temperature dependency of the electron mobility in InAs/GaSb-based T2SL may indicate stronger electron-phonon coupling than in the bulk material. The increased mobility near 50 K could be attributed to a longer scattering time or a lower electron-effective mass at the CB edge.\\
\indent When the Hall scattering factor $r_{_H}$, deviates significantly from unity, it indicates that to derive the electron drift mobility from the experimentally calculated Hall mobility data, the Hall scattering factor must be precisely determined. Figure \ref{Hall_factor} shows the predicted values of the Hall scattering factor against the temperature at $B=0.69\:\:T$ for $N_{D}=9\times10^{17} \:{cm^{-3}}$, while Fig. \ref{Hall_factor2} depicts the Hall scattering factor as a function of temperature and the carrier concentration at $B=0.69\:\:T$.\\
\indent To the best of our knowledge, calculations of the Hall scattering factor in such SLs have not been performed yet in earlier works. The contribution of various scattering mechanisms decides the Hall scattering factor's value. Figures \ref{Hall_factor} and \ref{Hall_factor2} indicate that the value of $r_{_H}$ at low temperatures deviates significantly from unity, while many researchers use one as an ideal value for a variety of calculations and studies, which is not accurate. The carrier concentration and the drift mobility may both be overestimated and underestimated when the Hall scattering factor is used as unity. The Hall scattering factor, in our calculation, fluctuates between the values as low as 0.3 at low temperature and electron concentration, and as high as 1.48 and even more at high temperature and electron concentration as shown in Fig. \ref{Hall_factor2}. Therefore, it is worth pointing out that, while evaluating the carrier concentration and the drift mobility in such SLs, one must use caution.\\
\indent In this work, we calculate the precise values of the Hall scattering factor and show that for a doping value of $N_{D}=9\times10^{17}~{cm^{-3}}$, the computed values of $r_H$ are 0.914, 0.952 and 1.01 at temperatures of 77 K, 150 K and 190 K, respectively, as also depicted in Fig. \ref{Hall_factor}. At higher temperatures, the value of the Hall scattering factor is more than unity, indicating that the drift mobility is lower than the Hall mobility, implying that the phonon-assisted scattering mechanisms are substantial and diminish the drift mobility. As shown in Fig. \ref{Hall_factor2}, at temperatures of 30 K and 77 K, the Hall scattering factor is equal to 0.335 \& 0.638 for lower doping concentrations of $N_{D}=1\times10^{12}~{cm^{-3}}$ and it is equal to 0.369 \& 0.691 with slightly higher doping concentrations of $N_{D}=5\times10^{15}~{cm^{-3}}$ which signifies that the Hall scattering factor increases as the temperature and electron concentrations rise, but as we increase the carrier concentration beyond $3\times10^{17}~{cm^{-3}}$, the Hall scattering factor starts decreasing. The higher electron concentration causes a rapid variation in the Hall factor.
\section{Conclusion}
\label{conclu}
In this paper, we developed the Rode algorithm on the BTE in conjunction with the $\bf k.p$ band structure and the EFA for a detailed computation of the carrier mobility and conductivity, in order to primarily unravel two crucial insights. First, the significance of both elastic and inelastic scattering mechanisms, particularly the influence of the IRS and POP scattering mechanisms in technologically relevant SL structures. Second, the structure specific Hall mobility and Hall scattering factor, which reveals that temperature and carrier concentrations significantly affect the Hall scattering factor, which deviates significantly from unity, i.e., from 0.3 to about 1.48, even for small magnetic fields. This reinforces the caution that should be exercised when employing the Hall scattering factor in experimental estimations of drift mobilities and carrier concentrations. Our research offers a comprehensive microscopic understanding of carrier dynamics in such technologically relevant SLs. Our model also provides highly accurate and precise transport parameters beyond the RTA and hence paves the way to develop physics based device modules for MWIR photodetectors.
\section*{Acknowledgments}
The authors acknowledge funding from ISRO under the ISRO-IIT Bombay Space Technology Cell.
\bibliography{reference}

\begin{thebibliography}{78}
\expandafter\ifx\csname natexlab\endcsname\relax\def\natexlab#1{#1}\fi
\expandafter\ifx\csname bibnamefont\endcsname\relax
  \def\bibnamefont#1{#1}\fi
\expandafter\ifx\csname bibfnamefont\endcsname\relax
  \def\bibfnamefont#1{#1}\fi
\expandafter\ifx\csname citenamefont\endcsname\relax
  \def\citenamefont#1{#1}\fi
\expandafter\ifx\csname url\endcsname\relax
  \def\url#1{\texttt{#1}}\fi
\expandafter\ifx\csname urlprefix\endcsname\relax\def\urlprefix{URL }\fi
\providecommand{\bibinfo}[2]{#2}
\providecommand{\eprint}[2][]{\url{#2}}

\bibitem[{\citenamefont{Smith and Mailhiot}(1987)}]{smith1987proposal}
\bibinfo{author}{\bibfnamefont{D.}~\bibnamefont{Smith}} \bibnamefont{and}
  \bibinfo{author}{\bibfnamefont{C.}~\bibnamefont{Mailhiot}},
  \bibinfo{journal}{Journal of Applied Physics} \textbf{\bibinfo{volume}{62}},
  \bibinfo{pages}{2545} (\bibinfo{year}{1987}).

\bibitem[{\citenamefont{Rogalski et~al.}(2017)\citenamefont{Rogalski,
  Martyniuk, and Kopytko}}]{rogalski2017inas}
\bibinfo{author}{\bibfnamefont{A.}~\bibnamefont{Rogalski}},
  \bibinfo{author}{\bibfnamefont{P.}~\bibnamefont{Martyniuk}},
  \bibnamefont{and} \bibinfo{author}{\bibfnamefont{M.}~\bibnamefont{Kopytko}},
  \bibinfo{journal}{Applied physics reviews} \textbf{\bibinfo{volume}{4}},
  \bibinfo{pages}{031304} (\bibinfo{year}{2017}).

\bibitem[{\citenamefont{Mukherjee et~al.}(2021)\citenamefont{Mukherjee, Singh,
  Bodhankar, and Muralidharan}}]{mukherjee2021carrier}
\bibinfo{author}{\bibfnamefont{S.}~\bibnamefont{Mukherjee}},
  \bibinfo{author}{\bibfnamefont{A.}~\bibnamefont{Singh}},
  \bibinfo{author}{\bibfnamefont{A.}~\bibnamefont{Bodhankar}},
  \bibnamefont{and}
  \bibinfo{author}{\bibfnamefont{B.}~\bibnamefont{Muralidharan}},
  \bibinfo{journal}{Journal of Physics D: Applied Physics}
  \textbf{\bibinfo{volume}{54}}, \bibinfo{pages}{345104}
  (\bibinfo{year}{2021}).

\bibitem[{\citenamefont{Rogalski}(2000)}]{rogalski2000infrared}
\bibinfo{author}{\bibfnamefont{A.}~\bibnamefont{Rogalski}},
  \emph{\bibinfo{title}{Infrared detectors}} (\bibinfo{publisher}{CRC press},
  \bibinfo{year}{2000}).

\bibitem[{\citenamefont{Dehzangi et~al.}(2021)\citenamefont{Dehzangi, Li, and
  Razeghi}}]{dehzangi2021band}
\bibinfo{author}{\bibfnamefont{A.}~\bibnamefont{Dehzangi}},
  \bibinfo{author}{\bibfnamefont{J.}~\bibnamefont{Li}}, \bibnamefont{and}
  \bibinfo{author}{\bibfnamefont{M.}~\bibnamefont{Razeghi}},
  \bibinfo{journal}{Light: Science \& Applications}
  \textbf{\bibinfo{volume}{10}}, \bibinfo{pages}{1} (\bibinfo{year}{2021}).

\bibitem[{\citenamefont{Rogalski}(2003)}]{rogalski2003infrared}
\bibinfo{author}{\bibfnamefont{A.}~\bibnamefont{Rogalski}},
  \bibinfo{journal}{Progress in quantum electronics}
  \textbf{\bibinfo{volume}{27}}, \bibinfo{pages}{59} (\bibinfo{year}{2003}).

\bibitem[{\citenamefont{Le~Thi et~al.}(2019)\citenamefont{Le~Thi, Kamakura, and
  Mori}}]{le2019simulation}
\bibinfo{author}{\bibfnamefont{Y.}~\bibnamefont{Le~Thi}},
  \bibinfo{author}{\bibfnamefont{Y.}~\bibnamefont{Kamakura}}, \bibnamefont{and}
  \bibinfo{author}{\bibfnamefont{N.}~\bibnamefont{Mori}},
  \bibinfo{journal}{Japanese Journal of Applied Physics}
  \textbf{\bibinfo{volume}{58}}, \bibinfo{pages}{044002}
  (\bibinfo{year}{2019}).

\bibitem[{\citenamefont{Klipstein et~al.}(2021)\citenamefont{Klipstein, Benny,
  Cohen, Fraenkel, Fraenkel, Gliksman, Glozman, Hirsch, Klin, Langof
  et~al.}}]{klipstein2021type}
\bibinfo{author}{\bibfnamefont{P.}~\bibnamefont{Klipstein}},
  \bibinfo{author}{\bibfnamefont{Y.}~\bibnamefont{Benny}},
  \bibinfo{author}{\bibfnamefont{Y.}~\bibnamefont{Cohen}},
  \bibinfo{author}{\bibfnamefont{N.}~\bibnamefont{Fraenkel}},
  \bibinfo{author}{\bibfnamefont{R.}~\bibnamefont{Fraenkel}},
  \bibinfo{author}{\bibfnamefont{S.}~\bibnamefont{Gliksman}},
  \bibinfo{author}{\bibfnamefont{A.}~\bibnamefont{Glozman}},
  \bibinfo{author}{\bibfnamefont{I.}~\bibnamefont{Hirsch}},
  \bibinfo{author}{\bibfnamefont{O.}~\bibnamefont{Klin}},
  \bibinfo{author}{\bibfnamefont{L.}~\bibnamefont{Langof}},
  \bibnamefont{et~al.}, in \emph{\bibinfo{booktitle}{Infrared Technology and
  Applications XLVII}} (\bibinfo{organization}{SPIE}, \bibinfo{year}{2021}),
  vol. \bibinfo{volume}{11741}, pp. \bibinfo{pages}{102--112}.

\bibitem[{\citenamefont{Wr{\'o}bel et~al.}(2012)\citenamefont{Wr{\'o}bel,
  Martyniuk, Plis, Madejczyk, Gawron, Krishna, and Rogalski}}]{wrobel2012dark}
\bibinfo{author}{\bibfnamefont{J.}~\bibnamefont{Wr{\'o}bel}},
  \bibinfo{author}{\bibfnamefont{P.}~\bibnamefont{Martyniuk}},
  \bibinfo{author}{\bibfnamefont{E.}~\bibnamefont{Plis}},
  \bibinfo{author}{\bibfnamefont{P.}~\bibnamefont{Madejczyk}},
  \bibinfo{author}{\bibfnamefont{W.}~\bibnamefont{Gawron}},
  \bibinfo{author}{\bibfnamefont{S.}~\bibnamefont{Krishna}}, \bibnamefont{and}
  \bibinfo{author}{\bibfnamefont{A.}~\bibnamefont{Rogalski}}, in
  \emph{\bibinfo{booktitle}{Infrared Technology and Applications XXXVIII}}
  (\bibinfo{organization}{SPIE}, \bibinfo{year}{2012}), vol.
  \bibinfo{volume}{8353}, pp. \bibinfo{pages}{412--420}.

\bibitem[{\citenamefont{Gautam et~al.}(2010)\citenamefont{Gautam, Kim, Kutty,
  Plis, Dawson, and Krishna}}]{gautam2010performance}
\bibinfo{author}{\bibfnamefont{N.}~\bibnamefont{Gautam}},
  \bibinfo{author}{\bibfnamefont{H.}~\bibnamefont{Kim}},
  \bibinfo{author}{\bibfnamefont{M.}~\bibnamefont{Kutty}},
  \bibinfo{author}{\bibfnamefont{E.}~\bibnamefont{Plis}},
  \bibinfo{author}{\bibfnamefont{L.}~\bibnamefont{Dawson}}, \bibnamefont{and}
  \bibinfo{author}{\bibfnamefont{S.}~\bibnamefont{Krishna}},
  \bibinfo{journal}{Applied Physics Letters} \textbf{\bibinfo{volume}{96}},
  \bibinfo{pages}{231107} (\bibinfo{year}{2010}).

\bibitem[{\citenamefont{Sai-Halasz et~al.}(1978)\citenamefont{Sai-Halasz,
  Esaki, and Harrison}}]{SaiHalasz}
\bibinfo{author}{\bibfnamefont{G.~A.} \bibnamefont{Sai-Halasz}},
  \bibinfo{author}{\bibfnamefont{L.}~\bibnamefont{Esaki}}, \bibnamefont{and}
  \bibinfo{author}{\bibfnamefont{W.~A.} \bibnamefont{Harrison}},
  \bibinfo{journal}{Phys. Rev. B} \textbf{\bibinfo{volume}{18}},
  \bibinfo{pages}{2812} (\bibinfo{year}{1978}).

\bibitem[{\citenamefont{Manyk et~al.}(2018)\citenamefont{Manyk, Michalczewski,
  Murawski, Grodecki, Rutkowski, and Martyniuk}}]{manyk2018electronic}
\bibinfo{author}{\bibfnamefont{T.}~\bibnamefont{Manyk}},
  \bibinfo{author}{\bibfnamefont{K.}~\bibnamefont{Michalczewski}},
  \bibinfo{author}{\bibfnamefont{K.}~\bibnamefont{Murawski}},
  \bibinfo{author}{\bibfnamefont{K.}~\bibnamefont{Grodecki}},
  \bibinfo{author}{\bibfnamefont{J.}~\bibnamefont{Rutkowski}},
  \bibnamefont{and}
  \bibinfo{author}{\bibfnamefont{P.}~\bibnamefont{Martyniuk}},
  \bibinfo{journal}{Results in Physics} \textbf{\bibinfo{volume}{11}},
  \bibinfo{pages}{1119} (\bibinfo{year}{2018}).

\bibitem[{\citenamefont{Rogalski et~al.}(2019)\citenamefont{Rogalski,
  Martyniuk, and Kopytko}}]{rogalski2019type}
\bibinfo{author}{\bibfnamefont{A.}~\bibnamefont{Rogalski}},
  \bibinfo{author}{\bibfnamefont{P.}~\bibnamefont{Martyniuk}},
  \bibnamefont{and} \bibinfo{author}{\bibfnamefont{M.}~\bibnamefont{Kopytko}},
  \bibinfo{journal}{Progress in Quantum Electronics}
  \textbf{\bibinfo{volume}{68}}, \bibinfo{pages}{100228}
  (\bibinfo{year}{2019}).

\bibitem[{\citenamefont{Chow et~al.}(1991)\citenamefont{Chow, Miles, Schulman,
  Collins, and McGill}}]{chow1991type}
\bibinfo{author}{\bibfnamefont{D.}~\bibnamefont{Chow}},
  \bibinfo{author}{\bibfnamefont{R.}~\bibnamefont{Miles}},
  \bibinfo{author}{\bibfnamefont{J.}~\bibnamefont{Schulman}},
  \bibinfo{author}{\bibfnamefont{D.}~\bibnamefont{Collins}}, \bibnamefont{and}
  \bibinfo{author}{\bibfnamefont{T.}~\bibnamefont{McGill}},
  \bibinfo{journal}{Semiconductor Science and Technology}
  \textbf{\bibinfo{volume}{6}}, \bibinfo{pages}{C47} (\bibinfo{year}{1991}).

\bibitem[{\citenamefont{Plis}(2014)}]{plis2014inas}
\bibinfo{author}{\bibfnamefont{E.~A.} \bibnamefont{Plis}},
  \bibinfo{journal}{Advances in Electronics} \textbf{\bibinfo{volume}{2014}}
  (\bibinfo{year}{2014}).

\bibitem[{\citenamefont{Martyniuk et~al.}(2014)\citenamefont{Martyniuk,
  Antoszewski, Martyniuk, Faraone, and Rogalski}}]{martyniuk2014new}
\bibinfo{author}{\bibfnamefont{P.}~\bibnamefont{Martyniuk}},
  \bibinfo{author}{\bibfnamefont{J.}~\bibnamefont{Antoszewski}},
  \bibinfo{author}{\bibfnamefont{M.}~\bibnamefont{Martyniuk}},
  \bibinfo{author}{\bibfnamefont{L.}~\bibnamefont{Faraone}}, \bibnamefont{and}
  \bibinfo{author}{\bibfnamefont{A.}~\bibnamefont{Rogalski}},
  \bibinfo{journal}{Applied Physics Reviews} \textbf{\bibinfo{volume}{1}},
  \bibinfo{pages}{041102} (\bibinfo{year}{2014}).

\bibitem[{\citenamefont{Safa et~al.}(2013)\citenamefont{Safa, Asgari, and
  Faraone}}]{safa2013study}
\bibinfo{author}{\bibfnamefont{S.}~\bibnamefont{Safa}},
  \bibinfo{author}{\bibfnamefont{A.}~\bibnamefont{Asgari}}, \bibnamefont{and}
  \bibinfo{author}{\bibfnamefont{L.}~\bibnamefont{Faraone}},
  \bibinfo{journal}{Journal of Applied Physics} \textbf{\bibinfo{volume}{114}},
  \bibinfo{pages}{053712} (\bibinfo{year}{2013}).

\bibitem[{\citenamefont{Safa and Asgari}(2015{\natexlab{a}})}]{safa2015role}
\bibinfo{author}{\bibfnamefont{S.}~\bibnamefont{Safa}} \bibnamefont{and}
  \bibinfo{author}{\bibfnamefont{A.}~\bibnamefont{Asgari}},
  \bibinfo{journal}{arXiv preprint arXiv:1502.02449}
  (\bibinfo{year}{2015}{\natexlab{a}}).

\bibitem[{\citenamefont{Safa and Asgari}(2015{\natexlab{b}})}]{safa2015study}
\bibinfo{author}{\bibfnamefont{S.}~\bibnamefont{Safa}} \bibnamefont{and}
  \bibinfo{author}{\bibfnamefont{A.}~\bibnamefont{Asgari}},
  \bibinfo{journal}{arXiv preprint arXiv:1502.01453}
  (\bibinfo{year}{2015}{\natexlab{b}}).

\bibitem[{\citenamefont{Safa and
  Asgari}(2015{\natexlab{c}})}]{safa2015vertical}
\bibinfo{author}{\bibfnamefont{S.}~\bibnamefont{Safa}} \bibnamefont{and}
  \bibinfo{author}{\bibfnamefont{A.}~\bibnamefont{Asgari}},
  \bibinfo{journal}{arXiv preprint arXiv:1504.02871}
  (\bibinfo{year}{2015}{\natexlab{c}}).

\bibitem[{\citenamefont{Szmulowicz et~al.}(2011)\citenamefont{Szmulowicz,
  Haugan, Elhamri, and Brown}}]{szmulowicz2011calculation}
\bibinfo{author}{\bibfnamefont{F.}~\bibnamefont{Szmulowicz}},
  \bibinfo{author}{\bibfnamefont{H.}~\bibnamefont{Haugan}},
  \bibinfo{author}{\bibfnamefont{S.}~\bibnamefont{Elhamri}}, \bibnamefont{and}
  \bibinfo{author}{\bibfnamefont{G.}~\bibnamefont{Brown}},
  \bibinfo{journal}{Physical Review B} \textbf{\bibinfo{volume}{84}},
  \bibinfo{pages}{155307} (\bibinfo{year}{2011}).

\bibitem[{\citenamefont{Szmulowicz and
  Brown}(2011)}]{szmulowicz2011calculation2}
\bibinfo{author}{\bibfnamefont{F.}~\bibnamefont{Szmulowicz}} \bibnamefont{and}
  \bibinfo{author}{\bibfnamefont{G.}~\bibnamefont{Brown}},
  \bibinfo{journal}{Applied Physics Letters} \textbf{\bibinfo{volume}{98}},
  \bibinfo{pages}{182105} (\bibinfo{year}{2011}).

\bibitem[{\citenamefont{Szmulowicz and
  Brown}(2013)}]{szmulowicz2013calculation}
\bibinfo{author}{\bibfnamefont{F.}~\bibnamefont{Szmulowicz}} \bibnamefont{and}
  \bibinfo{author}{\bibfnamefont{G.}~\bibnamefont{Brown}},
  \bibinfo{journal}{Journal of Applied Physics} \textbf{\bibinfo{volume}{113}},
  \bibinfo{pages}{014302} (\bibinfo{year}{2013}).

\bibitem[{\citenamefont{Bastard}(1981)}]{bastard1981superlattice}
\bibinfo{author}{\bibfnamefont{G.}~\bibnamefont{Bastard}},
  \bibinfo{journal}{Physical Review B} \textbf{\bibinfo{volume}{24}},
  \bibinfo{pages}{5693} (\bibinfo{year}{1981}).

\bibitem[{\citenamefont{Conwell and Weisskopf}(1950)}]{conwell1950theory}
\bibinfo{author}{\bibfnamefont{E.}~\bibnamefont{Conwell}} \bibnamefont{and}
  \bibinfo{author}{\bibfnamefont{V.}~\bibnamefont{Weisskopf}},
  \bibinfo{journal}{Physical review} \textbf{\bibinfo{volume}{77}},
  \bibinfo{pages}{388} (\bibinfo{year}{1950}).

\bibitem[{\citenamefont{Zook}(1964)}]{zook1964piezoelectric}
\bibinfo{author}{\bibfnamefont{J.~D.} \bibnamefont{Zook}},
  \bibinfo{journal}{Physical Review} \textbf{\bibinfo{volume}{136}},
  \bibinfo{pages}{A869} (\bibinfo{year}{1964}).

\bibitem[{\citenamefont{Kaasbjerg et~al.}(2013)\citenamefont{Kaasbjerg,
  Thygesen, and Jauho}}]{kaasbjerg2013acoustic}
\bibinfo{author}{\bibfnamefont{K.}~\bibnamefont{Kaasbjerg}},
  \bibinfo{author}{\bibfnamefont{K.~S.} \bibnamefont{Thygesen}},
  \bibnamefont{and} \bibinfo{author}{\bibfnamefont{A.-P.} \bibnamefont{Jauho}},
  \bibinfo{journal}{Physical Review B} \textbf{\bibinfo{volume}{87}},
  \bibinfo{pages}{235312} (\bibinfo{year}{2013}).

\bibitem[{\citenamefont{Tsai et~al.}(2020)\citenamefont{Tsai, Michalczewski,
  Martyniuk, Wu, and Wu}}]{tsai2020application}
\bibinfo{author}{\bibfnamefont{T.-Y.} \bibnamefont{Tsai}},
  \bibinfo{author}{\bibfnamefont{K.}~\bibnamefont{Michalczewski}},
  \bibinfo{author}{\bibfnamefont{P.}~\bibnamefont{Martyniuk}},
  \bibinfo{author}{\bibfnamefont{C.-H.} \bibnamefont{Wu}}, \bibnamefont{and}
  \bibinfo{author}{\bibfnamefont{Y.-R.} \bibnamefont{Wu}},
  \bibinfo{journal}{Journal of Applied Physics} \textbf{\bibinfo{volume}{127}},
  \bibinfo{pages}{033104} (\bibinfo{year}{2020}).

\bibitem[{\citenamefont{Wataya et~al.}(1989)\citenamefont{Wataya, Sawaki, Goto,
  Akasaki, Kano, and Hashimoto}}]{wataya1989interface}
\bibinfo{author}{\bibfnamefont{M.}~\bibnamefont{Wataya}},
  \bibinfo{author}{\bibfnamefont{N.}~\bibnamefont{Sawaki}},
  \bibinfo{author}{\bibfnamefont{H.}~\bibnamefont{Goto}},
  \bibinfo{author}{\bibfnamefont{I.}~\bibnamefont{Akasaki}},
  \bibinfo{author}{\bibfnamefont{H.}~\bibnamefont{Kano}}, \bibnamefont{and}
  \bibinfo{author}{\bibfnamefont{M.}~\bibnamefont{Hashimoto}},
  \bibinfo{journal}{Japanese Journal of Applied Physics}
  \textbf{\bibinfo{volume}{28}}, \bibinfo{pages}{1934} (\bibinfo{year}{1989}).

\bibitem[{\citenamefont{Dharssi et~al.}(1991)\citenamefont{Dharssi, Butcher,
  and Warren}}]{dharssi1991mobility}
\bibinfo{author}{\bibfnamefont{I.}~\bibnamefont{Dharssi}},
  \bibinfo{author}{\bibfnamefont{P.}~\bibnamefont{Butcher}}, \bibnamefont{and}
  \bibinfo{author}{\bibfnamefont{G.}~\bibnamefont{Warren}},
  \bibinfo{journal}{Superlattices and microstructures}
  \textbf{\bibinfo{volume}{9}}, \bibinfo{pages}{335} (\bibinfo{year}{1991}).

\bibitem[{\citenamefont{Dharssi and Butcher}(1990)}]{dharssi1990interface}
\bibinfo{author}{\bibfnamefont{I.}~\bibnamefont{Dharssi}} \bibnamefont{and}
  \bibinfo{author}{\bibfnamefont{P.}~\bibnamefont{Butcher}},
  \bibinfo{journal}{Journal of Physics: condensed matter}
  \textbf{\bibinfo{volume}{2}}, \bibinfo{pages}{4629} (\bibinfo{year}{1990}).

\bibitem[{\citenamefont{Rode}(1970)}]{rode1970electron}
\bibinfo{author}{\bibfnamefont{D.}~\bibnamefont{Rode}},
  \bibinfo{journal}{Physical Review B} \textbf{\bibinfo{volume}{2}},
  \bibinfo{pages}{1012} (\bibinfo{year}{1970}).

\bibitem[{\citenamefont{Rode}(1973)}]{rode1973theory}
\bibinfo{author}{\bibfnamefont{D.}~\bibnamefont{Rode}},
  \bibinfo{journal}{physica status solidi (b)} \textbf{\bibinfo{volume}{55}},
  \bibinfo{pages}{687} (\bibinfo{year}{1973}).

\bibitem[{\citenamefont{Rode}(1975)}]{rode1975low}
\bibinfo{author}{\bibfnamefont{D.}~\bibnamefont{Rode}}, in
  \emph{\bibinfo{booktitle}{Semiconductors and semimetals}}
  (\bibinfo{publisher}{Elsevier}, \bibinfo{year}{1975}),
  vol.~\bibinfo{volume}{10}, pp. \bibinfo{pages}{1--89}.

\bibitem[{\citenamefont{Ashcroft et~al.}()\citenamefont{Ashcroft, Mermin
  et~al.}}]{ashcroft1976solid}
\bibinfo{author}{\bibfnamefont{N.~W.} \bibnamefont{Ashcroft}},
  \bibinfo{author}{\bibfnamefont{N.~D.} \bibnamefont{Mermin}},
  \bibnamefont{et~al.}, \emph{\bibinfo{title}{Solid state physics}}.

\bibitem[{\citenamefont{Mermin}(1970)}]{mermin1970lindhard}
\bibinfo{author}{\bibfnamefont{N.~D.} \bibnamefont{Mermin}},
  \bibinfo{journal}{Physical Review B} \textbf{\bibinfo{volume}{1}},
  \bibinfo{pages}{2362} (\bibinfo{year}{1970}).

\bibitem[{\citenamefont{Livneh et~al.}(2012)\citenamefont{Livneh, Klipstein,
  Klin, Snapi, Grossman, Glozman, and Weiss}}]{livneh2012k}
\bibinfo{author}{\bibfnamefont{Y.}~\bibnamefont{Livneh}},
  \bibinfo{author}{\bibfnamefont{P.}~\bibnamefont{Klipstein}},
  \bibinfo{author}{\bibfnamefont{O.}~\bibnamefont{Klin}},
  \bibinfo{author}{\bibfnamefont{N.}~\bibnamefont{Snapi}},
  \bibinfo{author}{\bibfnamefont{S.}~\bibnamefont{Grossman}},
  \bibinfo{author}{\bibfnamefont{A.}~\bibnamefont{Glozman}}, \bibnamefont{and}
  \bibinfo{author}{\bibfnamefont{E.}~\bibnamefont{Weiss}},
  \bibinfo{journal}{Physical Review B} \textbf{\bibinfo{volume}{86}},
  \bibinfo{pages}{235311} (\bibinfo{year}{2012}).

\bibitem[{\citenamefont{Klipstein}(2010)}]{klipstein2010operator}
\bibinfo{author}{\bibfnamefont{P.}~\bibnamefont{Klipstein}},
  \bibinfo{journal}{Physical Review B} \textbf{\bibinfo{volume}{81}},
  \bibinfo{pages}{235314} (\bibinfo{year}{2010}).

\bibitem[{\citenamefont{Ricciardi et~al.}(2020)\citenamefont{Ricciardi,
  Della~Rocca, and Benfante}}]{ricciardi2020ingaas}
\bibinfo{author}{\bibfnamefont{C.}~\bibnamefont{Ricciardi}},
  \bibinfo{author}{\bibfnamefont{M.~L.} \bibnamefont{Della~Rocca}},
  \bibnamefont{and} \bibinfo{author}{\bibfnamefont{M.}~\bibnamefont{Benfante}}
  (\bibinfo{year}{2020}).

\bibitem[{\citenamefont{Qiao et~al.}(2012)\citenamefont{Qiao, Mou, and
  Chuang}}]{qiao2012electronic}
\bibinfo{author}{\bibfnamefont{P.-F.} \bibnamefont{Qiao}},
  \bibinfo{author}{\bibfnamefont{S.}~\bibnamefont{Mou}}, \bibnamefont{and}
  \bibinfo{author}{\bibfnamefont{S.~L.} \bibnamefont{Chuang}},
  \bibinfo{journal}{Optics express} \textbf{\bibinfo{volume}{20}},
  \bibinfo{pages}{2319} (\bibinfo{year}{2012}).

\bibitem[{\citenamefont{Klipstein et~al.}(2013)\citenamefont{Klipstein, Livneh,
  Klin, Grossman, Snapi, Glozman, and Weiss}}]{klipstein2013k}
\bibinfo{author}{\bibfnamefont{P.}~\bibnamefont{Klipstein}},
  \bibinfo{author}{\bibfnamefont{Y.}~\bibnamefont{Livneh}},
  \bibinfo{author}{\bibfnamefont{O.}~\bibnamefont{Klin}},
  \bibinfo{author}{\bibfnamefont{S.}~\bibnamefont{Grossman}},
  \bibinfo{author}{\bibfnamefont{N.}~\bibnamefont{Snapi}},
  \bibinfo{author}{\bibfnamefont{A.}~\bibnamefont{Glozman}}, \bibnamefont{and}
  \bibinfo{author}{\bibfnamefont{E.}~\bibnamefont{Weiss}},
  \bibinfo{journal}{Infrared Physics \& Technology}
  \textbf{\bibinfo{volume}{59}}, \bibinfo{pages}{53} (\bibinfo{year}{2013}).

\bibitem[{\citenamefont{Aspnes and Studna}(1983)}]{aspnes1983dielectric}
\bibinfo{author}{\bibfnamefont{D.~E.} \bibnamefont{Aspnes}} \bibnamefont{and}
  \bibinfo{author}{\bibfnamefont{A.}~\bibnamefont{Studna}},
  \bibinfo{journal}{Physical review B} \textbf{\bibinfo{volume}{27}},
  \bibinfo{pages}{985} (\bibinfo{year}{1983}).

\bibitem[{\citenamefont{Datta}(2005)}]{datta2005quantum}
\bibinfo{author}{\bibfnamefont{S.}~\bibnamefont{Datta}},
  \emph{\bibinfo{title}{Quantum transport: atom to transistor}}
  (\bibinfo{publisher}{Cambridge university press}, \bibinfo{year}{2005}).

\bibitem[{\citenamefont{Garwood et~al.}(2017)\citenamefont{Garwood, Modine, and
  Krishna}}]{garwood2017electronic}
\bibinfo{author}{\bibfnamefont{T.}~\bibnamefont{Garwood}},
  \bibinfo{author}{\bibfnamefont{N.~A.} \bibnamefont{Modine}},
  \bibnamefont{and} \bibinfo{author}{\bibfnamefont{S.}~\bibnamefont{Krishna}},
  \bibinfo{journal}{Infrared Physics \& Technology}
  \textbf{\bibinfo{volume}{81}}, \bibinfo{pages}{27} (\bibinfo{year}{2017}).

\bibitem[{\citenamefont{Wei and Razeghi}(2004)}]{wei2004modeling}
\bibinfo{author}{\bibfnamefont{Y.}~\bibnamefont{Wei}} \bibnamefont{and}
  \bibinfo{author}{\bibfnamefont{M.}~\bibnamefont{Razeghi}},
  \bibinfo{journal}{Physical Review B} \textbf{\bibinfo{volume}{69}},
  \bibinfo{pages}{085316} (\bibinfo{year}{2004}).

\bibitem[{\citenamefont{Nucho and Madhukar}(1978)}]{nucho1978tight}
\bibinfo{author}{\bibfnamefont{R.}~\bibnamefont{Nucho}} \bibnamefont{and}
  \bibinfo{author}{\bibfnamefont{A.}~\bibnamefont{Madhukar}},
  \bibinfo{journal}{Journal of Vacuum Science and Technology}
  \textbf{\bibinfo{volume}{15}}, \bibinfo{pages}{1530} (\bibinfo{year}{1978}).

\bibitem[{\citenamefont{Dente and Tilton}(1999)}]{dente1999pseudopotential}
\bibinfo{author}{\bibfnamefont{G.~C.} \bibnamefont{Dente}} \bibnamefont{and}
  \bibinfo{author}{\bibfnamefont{M.~L.} \bibnamefont{Tilton}},
  \bibinfo{journal}{Journal of Applied Physics} \textbf{\bibinfo{volume}{86}},
  \bibinfo{pages}{1420} (\bibinfo{year}{1999}).

\bibitem[{\citenamefont{Magri and Zunger}(2002)}]{magri2002effects}
\bibinfo{author}{\bibfnamefont{R.}~\bibnamefont{Magri}} \bibnamefont{and}
  \bibinfo{author}{\bibfnamefont{A.}~\bibnamefont{Zunger}},
  \bibinfo{journal}{Physical Review B} \textbf{\bibinfo{volume}{65}},
  \bibinfo{pages}{165302} (\bibinfo{year}{2002}).

\bibitem[{\citenamefont{Taghipour et~al.}(2018)\citenamefont{Taghipour,
  Shojaee, and Krishna}}]{taghipour2018many}
\bibinfo{author}{\bibfnamefont{Z.}~\bibnamefont{Taghipour}},
  \bibinfo{author}{\bibfnamefont{E.}~\bibnamefont{Shojaee}}, \bibnamefont{and}
  \bibinfo{author}{\bibfnamefont{S.}~\bibnamefont{Krishna}},
  \bibinfo{journal}{Journal of Physics: Condensed Matter}
  \textbf{\bibinfo{volume}{30}}, \bibinfo{pages}{325701}
  (\bibinfo{year}{2018}).

\bibitem[{\citenamefont{Bastard}(1982)}]{bastard1982theoretical}
\bibinfo{author}{\bibfnamefont{G.}~\bibnamefont{Bastard}},
  \bibinfo{journal}{Physical Review B} \textbf{\bibinfo{volume}{25}},
  \bibinfo{pages}{7584} (\bibinfo{year}{1982}).

\bibitem[{\citenamefont{Altarelli}(1983)}]{altarelli1983electronic}
\bibinfo{author}{\bibfnamefont{M.}~\bibnamefont{Altarelli}},
  \bibinfo{journal}{Physical review B} \textbf{\bibinfo{volume}{28}},
  \bibinfo{pages}{842} (\bibinfo{year}{1983}).

\bibitem[{\citenamefont{Kane}(1980)}]{kane1980band}
\bibinfo{author}{\bibfnamefont{E.}~\bibnamefont{Kane}}, in
  \emph{\bibinfo{booktitle}{Narrow Gap Semiconductors Physics and
  Applications}} (\bibinfo{publisher}{Springer}, \bibinfo{year}{1980}), pp.
  \bibinfo{pages}{13--31}.

\bibitem[{\citenamefont{Chuang}(2012)}]{chuang2012physics}
\bibinfo{author}{\bibfnamefont{S.~L.} \bibnamefont{Chuang}},
  \emph{\bibinfo{title}{Physics of photonic devices}} (\bibinfo{publisher}{John
  Wiley \& Sons}, \bibinfo{year}{2012}).

\bibitem[{\citenamefont{Jiang et~al.}(2014)\citenamefont{Jiang, Ma, Xu, and
  Song}}]{jiang2014finite}
\bibinfo{author}{\bibfnamefont{Y.}~\bibnamefont{Jiang}},
  \bibinfo{author}{\bibfnamefont{X.}~\bibnamefont{Ma}},
  \bibinfo{author}{\bibfnamefont{Y.}~\bibnamefont{Xu}}, \bibnamefont{and}
  \bibinfo{author}{\bibfnamefont{G.}~\bibnamefont{Song}},
  \bibinfo{journal}{Journal of Applied Physics} \textbf{\bibinfo{volume}{116}},
  \bibinfo{pages}{173702} (\bibinfo{year}{2014}).

\bibitem[{\citenamefont{Dhar et~al.}(2013)\citenamefont{Dhar, Dat, and
  Sood}}]{dhar2013advances}
\bibinfo{author}{\bibfnamefont{N.~K.} \bibnamefont{Dhar}},
  \bibinfo{author}{\bibfnamefont{R.}~\bibnamefont{Dat}}, \bibnamefont{and}
  \bibinfo{author}{\bibfnamefont{A.~K.} \bibnamefont{Sood}},
  \bibinfo{journal}{Optoelectronics-Advanced Materials and Devices}
  \textbf{\bibinfo{volume}{1}} (\bibinfo{year}{2013}).

\bibitem[{\citenamefont{Lundstrom}(2002)}]{lundstrom2002fundamentals}
\bibinfo{author}{\bibfnamefont{M.}~\bibnamefont{Lundstrom}},
  \emph{\bibinfo{title}{Fundamentals of carrier transport}}
  (\bibinfo{year}{2002}).

\bibitem[{\citenamefont{Ferry}(2016)}]{ferry2016semiconductor}
\bibinfo{author}{\bibfnamefont{D.~K.} \bibnamefont{Ferry}},
  \emph{\bibinfo{title}{Semiconductor transport}} (\bibinfo{publisher}{CRC
  Press}, \bibinfo{year}{2016}).

\bibitem[{\citenamefont{Singh}(2007)}]{singh2007electronic}
\bibinfo{author}{\bibfnamefont{J.}~\bibnamefont{Singh}},
  \emph{\bibinfo{title}{Electronic and optoelectronic properties of
  semiconductor structures}} (\bibinfo{publisher}{Cambridge University Press},
  \bibinfo{year}{2007}).

\bibitem[{\citenamefont{Pierret and Neudeck}(1987)}]{pierret1987advanced}
\bibinfo{author}{\bibfnamefont{R.~F.} \bibnamefont{Pierret}} \bibnamefont{and}
  \bibinfo{author}{\bibfnamefont{G.~W.} \bibnamefont{Neudeck}},
  \emph{\bibinfo{title}{Advanced semiconductor fundamentals}},
  vol.~\bibinfo{volume}{6} (\bibinfo{publisher}{Addison-Wesley Reading, MA},
  \bibinfo{year}{1987}).

\bibitem[{\citenamefont{Vasileska et~al.}(2017)\citenamefont{Vasileska,
  Goodnick, and Klimeck}}]{vasileska2017computational}
\bibinfo{author}{\bibfnamefont{D.}~\bibnamefont{Vasileska}},
  \bibinfo{author}{\bibfnamefont{S.~M.} \bibnamefont{Goodnick}},
  \bibnamefont{and} \bibinfo{author}{\bibfnamefont{G.}~\bibnamefont{Klimeck}},
  \emph{\bibinfo{title}{Computational Electronics: semiclassical and quantum
  device modeling and simulation}} (\bibinfo{publisher}{CRC press},
  \bibinfo{year}{2017}).

\bibitem[{\citenamefont{Chakrabarty et~al.}(2019)\citenamefont{Chakrabarty,
  Mandia, Muralidharan, Lee, and Bhattacharjee}}]{chakrabarty2019semi}
\bibinfo{author}{\bibfnamefont{S.}~\bibnamefont{Chakrabarty}},
  \bibinfo{author}{\bibfnamefont{A.~K.} \bibnamefont{Mandia}},
  \bibinfo{author}{\bibfnamefont{B.}~\bibnamefont{Muralidharan}},
  \bibinfo{author}{\bibfnamefont{S.~C.} \bibnamefont{Lee}}, \bibnamefont{and}
  \bibinfo{author}{\bibfnamefont{S.}~\bibnamefont{Bhattacharjee}},
  \bibinfo{journal}{Journal of Physics: Condensed Matter}
  \textbf{\bibinfo{volume}{32}}, \bibinfo{pages}{135704}
  (\bibinfo{year}{2019}).

\bibitem[{\citenamefont{Mandia et~al.}(2021)\citenamefont{Mandia, Muralidharan,
  Choi, Lee, and Bhattacharjee}}]{mandia2021ammcr}
\bibinfo{author}{\bibfnamefont{A.~K.} \bibnamefont{Mandia}},
  \bibinfo{author}{\bibfnamefont{B.}~\bibnamefont{Muralidharan}},
  \bibinfo{author}{\bibfnamefont{J.-H.} \bibnamefont{Choi}},
  \bibinfo{author}{\bibfnamefont{S.-C.} \bibnamefont{Lee}}, \bibnamefont{and}
  \bibinfo{author}{\bibfnamefont{S.}~\bibnamefont{Bhattacharjee}},
  \bibinfo{journal}{Computer Physics Communications}
  \textbf{\bibinfo{volume}{259}}, \bibinfo{pages}{107697}
  (\bibinfo{year}{2021}).

\bibitem[{\citenamefont{Ganose et~al.}(2021)\citenamefont{Ganose, Park,
  Faghaninia, Woods-Robinson, Persson, and Jain}}]{ganose2021efficient}
\bibinfo{author}{\bibfnamefont{A.~M.} \bibnamefont{Ganose}},
  \bibinfo{author}{\bibfnamefont{J.}~\bibnamefont{Park}},
  \bibinfo{author}{\bibfnamefont{A.}~\bibnamefont{Faghaninia}},
  \bibinfo{author}{\bibfnamefont{R.}~\bibnamefont{Woods-Robinson}},
  \bibinfo{author}{\bibfnamefont{K.~A.} \bibnamefont{Persson}},
  \bibnamefont{and} \bibinfo{author}{\bibfnamefont{A.}~\bibnamefont{Jain}},
  \bibinfo{journal}{Nature communications} \textbf{\bibinfo{volume}{12}},
  \bibinfo{pages}{1} (\bibinfo{year}{2021}).

\bibitem[{\citenamefont{Brooks}(1955)}]{brooks1955theory}
\bibinfo{author}{\bibfnamefont{H.}~\bibnamefont{Brooks}}, in
  \emph{\bibinfo{booktitle}{Advances in electronics and electron physics}}
  (\bibinfo{publisher}{Elsevier}, \bibinfo{year}{1955}),
  vol.~\bibinfo{volume}{7}, pp. \bibinfo{pages}{85--182}.

\bibitem[{\citenamefont{Faghaninia et~al.}(2015)\citenamefont{Faghaninia,
  Ager~III, and Lo}}]{faghaninia2015ab}
\bibinfo{author}{\bibfnamefont{A.}~\bibnamefont{Faghaninia}},
  \bibinfo{author}{\bibfnamefont{J.~W.} \bibnamefont{Ager~III}},
  \bibnamefont{and} \bibinfo{author}{\bibfnamefont{C.~S.} \bibnamefont{Lo}},
  \bibinfo{journal}{Physical Review B} \textbf{\bibinfo{volume}{91}},
  \bibinfo{pages}{235123} (\bibinfo{year}{2015}).

\bibitem[{\citenamefont{Kothari and Maldovan}(2017)}]{kothari2017phonon}
\bibinfo{author}{\bibfnamefont{K.}~\bibnamefont{Kothari}} \bibnamefont{and}
  \bibinfo{author}{\bibfnamefont{M.}~\bibnamefont{Maldovan}},
  \bibinfo{journal}{Scientific Reports} \textbf{\bibinfo{volume}{7}},
  \bibinfo{pages}{1} (\bibinfo{year}{2017}).

\bibitem[{\citenamefont{Sakaki et~al.}(1987)\citenamefont{Sakaki, Noda,
  Hirakawa, Tanaka, and Matsusue}}]{sakaki1987interface}
\bibinfo{author}{\bibfnamefont{H.}~\bibnamefont{Sakaki}},
  \bibinfo{author}{\bibfnamefont{T.}~\bibnamefont{Noda}},
  \bibinfo{author}{\bibfnamefont{K.}~\bibnamefont{Hirakawa}},
  \bibinfo{author}{\bibfnamefont{M.}~\bibnamefont{Tanaka}}, \bibnamefont{and}
  \bibinfo{author}{\bibfnamefont{T.}~\bibnamefont{Matsusue}},
  \bibinfo{journal}{Applied physics letters} \textbf{\bibinfo{volume}{51}},
  \bibinfo{pages}{1934} (\bibinfo{year}{1987}).

\bibitem[{\citenamefont{Gold}(1987)}]{gold1987electronic}
\bibinfo{author}{\bibfnamefont{A.}~\bibnamefont{Gold}},
  \bibinfo{journal}{Physical Review B} \textbf{\bibinfo{volume}{35}},
  \bibinfo{pages}{723} (\bibinfo{year}{1987}).

\bibitem[{\citenamefont{Sang et~al.}(2013)\citenamefont{Sang, Yang, Liu, Zhao,
  Liu, Gu, Wei, Liu, Zhu, and Wang}}]{sang2013dislocation}
\bibinfo{author}{\bibfnamefont{L.}~\bibnamefont{Sang}},
  \bibinfo{author}{\bibfnamefont{S.~Y.} \bibnamefont{Yang}},
  \bibinfo{author}{\bibfnamefont{G.~P.} \bibnamefont{Liu}},
  \bibinfo{author}{\bibfnamefont{G.~J.} \bibnamefont{Zhao}},
  \bibinfo{author}{\bibfnamefont{C.~B.} \bibnamefont{Liu}},
  \bibinfo{author}{\bibfnamefont{C.~Y.} \bibnamefont{Gu}},
  \bibinfo{author}{\bibfnamefont{H.~Y.} \bibnamefont{Wei}},
  \bibinfo{author}{\bibfnamefont{X.~L.} \bibnamefont{Liu}},
  \bibinfo{author}{\bibfnamefont{Q.~S.} \bibnamefont{Zhu}}, \bibnamefont{and}
  \bibinfo{author}{\bibfnamefont{Z.~G.} \bibnamefont{Wang}},
  \bibinfo{journal}{IEEE transactions on electron devices}
  \textbf{\bibinfo{volume}{60}}, \bibinfo{pages}{2077} (\bibinfo{year}{2013}).

\bibitem[{\citenamefont{Goodnick et~al.}(1985)\citenamefont{Goodnick, Ferry,
  Wilmsen, Liliental, Fathy, and Krivanek}}]{goodnick1985surface}
\bibinfo{author}{\bibfnamefont{S.}~\bibnamefont{Goodnick}},
  \bibinfo{author}{\bibfnamefont{D.}~\bibnamefont{Ferry}},
  \bibinfo{author}{\bibfnamefont{C.}~\bibnamefont{Wilmsen}},
  \bibinfo{author}{\bibfnamefont{Z.}~\bibnamefont{Liliental}},
  \bibinfo{author}{\bibfnamefont{D.}~\bibnamefont{Fathy}}, \bibnamefont{and}
  \bibinfo{author}{\bibfnamefont{O.}~\bibnamefont{Krivanek}},
  \bibinfo{journal}{Physical Review B} \textbf{\bibinfo{volume}{32}},
  \bibinfo{pages}{8171} (\bibinfo{year}{1985}).

\bibitem[{\citenamefont{Mandia et~al.}(2022)\citenamefont{Mandia, Koshi,
  Muralidharan, Lee, and Bhattacharjee}}]{mandia2022electrical}
\bibinfo{author}{\bibfnamefont{A.~K.} \bibnamefont{Mandia}},
  \bibinfo{author}{\bibfnamefont{N.~A.} \bibnamefont{Koshi}},
  \bibinfo{author}{\bibfnamefont{B.}~\bibnamefont{Muralidharan}},
  \bibinfo{author}{\bibfnamefont{S.~C.} \bibnamefont{Lee}}, \bibnamefont{and}
  \bibinfo{author}{\bibfnamefont{S.}~\bibnamefont{Bhattacharjee}},
  \bibinfo{journal}{Journal of Materials Chemistry C}  (\bibinfo{year}{2022}).

\bibitem[{\citenamefont{Becer et~al.}(2019)\citenamefont{Becer, Bennecer, and
  Sengouga}}]{becer2019modeling}
\bibinfo{author}{\bibfnamefont{Z.}~\bibnamefont{Becer}},
  \bibinfo{author}{\bibfnamefont{A.}~\bibnamefont{Bennecer}}, \bibnamefont{and}
  \bibinfo{author}{\bibfnamefont{N.}~\bibnamefont{Sengouga}},
  \bibinfo{journal}{Crystals} \textbf{\bibinfo{volume}{9}},
  \bibinfo{pages}{629} (\bibinfo{year}{2019}).

\bibitem[{\citenamefont{Vurgaftman et~al.}(2001)\citenamefont{Vurgaftman,
  Meyer, and Ram-Mohan}}]{vurgaftman2001band}
\bibinfo{author}{\bibfnamefont{I.}~\bibnamefont{Vurgaftman}},
  \bibinfo{author}{\bibfnamefont{J.~{\'a}.} \bibnamefont{Meyer}},
  \bibnamefont{and} \bibinfo{author}{\bibfnamefont{L.~{\'a}.}
  \bibnamefont{Ram-Mohan}}, \bibinfo{journal}{Journal of applied physics}
  \textbf{\bibinfo{volume}{89}}, \bibinfo{pages}{5815} (\bibinfo{year}{2001}).

\bibitem[{\citenamefont{Delmas et~al.}(2019)\citenamefont{Delmas, Liang, and
  Huffaker}}]{delmas2019comprehensive}
\bibinfo{author}{\bibfnamefont{M.}~\bibnamefont{Delmas}},
  \bibinfo{author}{\bibfnamefont{B.}~\bibnamefont{Liang}}, \bibnamefont{and}
  \bibinfo{author}{\bibfnamefont{D.~L.} \bibnamefont{Huffaker}}, in
  \emph{\bibinfo{booktitle}{Quantum Sensing and Nano Electronics and Photonics
  XVI}} (\bibinfo{organization}{International Society for Optics and
  Photonics}, \bibinfo{year}{2019}), vol. \bibinfo{volume}{10926}, p.
  \bibinfo{pages}{109260G}.

\bibitem[{\citenamefont{Mitra}(1963)}]{mitra1963phonon}
\bibinfo{author}{\bibfnamefont{S.}~\bibnamefont{Mitra}},
  \bibinfo{journal}{Physical Review} \textbf{\bibinfo{volume}{132}},
  \bibinfo{pages}{986} (\bibinfo{year}{1963}).

\bibitem[{\citenamefont{Lockwood et~al.}(2005)\citenamefont{Lockwood, Yu, and
  Rowell}}]{lockwood2005optical}
\bibinfo{author}{\bibfnamefont{D.}~\bibnamefont{Lockwood}},
  \bibinfo{author}{\bibfnamefont{G.}~\bibnamefont{Yu}}, \bibnamefont{and}
  \bibinfo{author}{\bibfnamefont{N.}~\bibnamefont{Rowell}},
  \bibinfo{journal}{Solid State Communications} \textbf{\bibinfo{volume}{136}},
  \bibinfo{pages}{404} (\bibinfo{year}{2005}).

\bibitem[{\citenamefont{Haugan et~al.}(2011)\citenamefont{Haugan, Brown,
  Szmulowicz, and Elhamri}}]{haugan2011design}
\bibinfo{author}{\bibfnamefont{H.}~\bibnamefont{Haugan}},
  \bibinfo{author}{\bibfnamefont{G.}~\bibnamefont{Brown}},
  \bibinfo{author}{\bibfnamefont{F.}~\bibnamefont{Szmulowicz}},
  \bibnamefont{and} \bibinfo{author}{\bibfnamefont{S.}~\bibnamefont{Elhamri}},
  in \emph{\bibinfo{booktitle}{AIP Conference Proceedings}}
  (\bibinfo{organization}{American Institute of Physics},
  \bibinfo{year}{2011}), vol. \bibinfo{volume}{1416}, pp.
  \bibinfo{pages}{155--157}.

\bibitem[{\citenamefont{Alchaar et~al.}(2019)\citenamefont{Alchaar, Rodriguez,
  H{\"o}glund, Naureen, and Christol}}]{alchaar2019characterization}
\bibinfo{author}{\bibfnamefont{R.}~\bibnamefont{Alchaar}},
  \bibinfo{author}{\bibfnamefont{J.-B.} \bibnamefont{Rodriguez}},
  \bibinfo{author}{\bibfnamefont{L.}~\bibnamefont{H{\"o}glund}},
  \bibinfo{author}{\bibfnamefont{S.}~\bibnamefont{Naureen}}, \bibnamefont{and}
  \bibinfo{author}{\bibfnamefont{P.}~\bibnamefont{Christol}},
  \bibinfo{journal}{AIP Advances} \textbf{\bibinfo{volume}{9}},
  \bibinfo{pages}{055012} (\bibinfo{year}{2019}).

\end{thebibliography}
\end{document}